\documentclass[final]{elsarticle} % This was 12pt

\usepackage{amssymb}
\usepackage{amsthm}
\usepackage{amsmath}
\usepackage{color}
\usepackage{mathtools}
\usepackage{graphicx}
\usepackage{lineno}
\usepackage{caption}
\usepackage{subcaption}
\usepackage{subfig}
\usepackage{tikz}
\usetikzlibrary{arrows}
\usepackage{hhline}
\usepackage{multirow}
\usepackage{pbox}
\usepackage{makecell}
\newcommand{\calN}{\mathcal{N}}
\newcommand{\calI}{\mathcal{I}}

\newcommand{\calT}{\mathcal{T}}
\newcommand{\calR}{\mathcal{R}}

\newcommand{\vecp}{\mathbf{p}}

\newcommand{\vecq}{\mathbf{q}}

\newcommand{\matR}{\mathbf{R}}

\newcommand{\qh}{\hat{q}}

\newlength{\Oldarrayrulewidth}
\newcommand{\Cline}[2]{%
  \noalign{\global\setlength{\Oldarrayrulewidth}{\arrayrulewidth}}%
  \noalign{\global\setlength{\arrayrulewidth}{#1}}\cline{#2}%
  \noalign{\global\setlength{\arrayrulewidth}{\Oldarrayrulewidth}}}
\usepackage{nomencl}
\usepackage{etoolbox}
\renewcommand\nomgroup[1]{%
  \item[\bfseries
  \ifstrequal{#1}{R}{$\;\;\;\;\;\;$Roman:}{\ifstrequal{#1}{M}{$\;\;\;\;\;\;$Mathematical:}{\ifstrequal{#1}{A}{$\;\;\;\;\;\;$Abbreviations:}{\ifstrequal{#1}{G}{$\;\;\;\;\;\;$Greek:}{\ifstrequal{#1}{S}{$\;\;\;\;\;\;$Subscripts:}{\ifstrequal{#1}{T}{$\;\;\;\;\;\;$Superscripts:}{}}}}}}%
]}

\makenomenclature

\journal{Journal of Computational Physics (DOI: 10.1016/j.jcp.2016.07.032)}

\begin{document}

\begin{frontmatter}



\title{Stability analysis of thermo-acoustic nonlinear eigenproblems in annular combustors. \\Part I. Sensitivity}


\author[label1,label2]{Luca Magri}
\author[label4]{Michael Bauerheim}
%\author[label3]{Franck Nicoud}
\author[label2]{Matthew P. Juniper}

\address[label1]{Center for Turbulence Research, Stanford University, CA, United States}
%\address[label3]{I3M UMR-CNSR 5149, University of Montpellier 2, France}
\address[label2]{Cambridge University Engineering Department, Cambridge, United Kingdom}
\address[label4]{CAPS Lab/, ETH Z\"urich, Switzerland}

\begin{abstract}
We present an adjoint-based method for the calculation of eigenvalue perturbations in nonlinear, degenerate and non self-adjoint eigenproblems. 
%The method is adjoint-based and enables us to calculate the sensitivities to perturbations to the system. 
This method is applied to a thermo-acoustic annular combustor network, the stability of which is governed by a nonlinear eigenproblem. 
We calculate the first- and second-order sensitivities of the growth rate and frequency to geometric, flow and flame parameters. Three different configurations are analysed. 
The benchmark sensitivities are obtained by finite difference, which involves solving the nonlinear eigenproblem at least as many times as the number of parameters. 
By solving only one adjoint eigenproblem, we obtain the sensitivities to any thermo-acoustic parameter, which match the finite-difference solutions at much lower computational cost.

\end{abstract}


\end{frontmatter}
%%%%%%%%%%%%%%%%%%%%%%%%%%%%%%%
%%%%%%%%%%%%%%%%%%%%%%%%%%%%%%%
% NOMENCLATURE
%%%%%%%%%%%%%%%%%%%%%%%%%%%%%%%
%%%%%%%%%%%%%%%%%%%%%%%%%%%%%%%
\renewcommand{\nomname}{Nomenclature}
%
%
%\nomenclature[A]{UQ}{Uncertainty Quantification}
%\nomenclature[A]{ASI}{Active Subspace Identification}
%\nomenclature[A]{RF}{Risk Factor}
\nomenclature[A]{AD}{Adjoint}
\nomenclature[A]{FD}{Finite difference}
%\nomenclature[A]{PDF}{Probability Density Function}
% %
% %
\nomenclature[G]{$\omega$}{Complex eigenvalue, $\omega_r+\mathrm{i}\omega_i$}
\nomenclature[G]{$\omega_r$}{Angular frequency}
\nomenclature[G]{$\omega_i$}{Growth rate}
%\nomenclature[G]{$\omega_i^{ASI}$}{Growth rate by surrogate models}
\nomenclature[G]{$\epsilon$}{Perturbation parameter}
% %
% %
\nomenclature[M]{$o$}{Little-o (Landau symbol)}
\nomenclature[M]{$\left\langle\cdot,\cdot\right\rangle$}{Inner product}
\nomenclature[M]{$\mathcal{O}$}{Big-O (Landau symbol)}
\nomenclature[M]{$\mathcal{N}$}{Operator representing the nonlinear eigenvalue problem}
% %
% %
\nomenclature[R]{$\mathbf{p}$}{Vector of thermo-acoustic parameters}
\nomenclature[R]{$\vecq$}{State vector}
%\nomenclature[R]{$M$}{Monte Carlo samples for uncertainty quantification}
%\nomenclature[R]{$M^{ASI}$}{Monte Carlo samples to build the covariance matrix}
\nomenclature[R]{$N$}{Eigenvalue geometric degeneracy}
%\nomenclature[R]{$\mathbf{e}_i$}{Degenerate vector}
\nomenclature[R]{$\mathrm{i}$}{Imaginary unit, $\mathrm{i}^2+1=0$}
% % 
% %
\nomenclature[S]{$0$}{Unperturbed}
\nomenclature[S]{$1$}{First-order perturbation}
\nomenclature[S]{$2$}{Second-order perturbation}
% % 
% %
\nomenclature[T]{$H$}{Hermitian}
%\nomenclature[T]{$T$}{Transpose}
\nomenclature[T]{$*$}{Complex conjugate}
\nomenclature[T]{$\hat{}$}{Eigenfunction}
\nomenclature[T]{$+$}{Adjoint}

\printnomenclature[10mm]
%%%%%%%%%%%%%%%%%%%%%%%%%%%%%%%
%%%%%%%%%%%%%%%%%%%%%%%%%%%%%%%
% INTRO 
%%%%%%%%%%%%%%%%%%%%%%%%%%%%%%%
%%%%%%%%%%%%%%%%%%%%%%%%%%%%%%%
\section{Introduction}\label{sec:intro}
Thermo-acoustic oscillations involve the interaction of heat release  and sound. In rocket and aircraft engines, heat release fluctuations can synchronize with the natural acoustic modes in the combustion chamber. This can cause loud vibrations that sometimes lead to catastrophic failure. It is one of the biggest and most persistent problems facing rocket and aircraft engine manufacturers \citep{Lieuwen2005}. 
%
%
%\subsection{Thermo-acoustics framework}
%

Many studies have demonstrated the ability of Large-Eddy Simulation (LES) to represent the flame dynamics \cite{Poinsot2013}.
However, even when LES simulations confirm that a combustor is unstable, they do not suggest how to control the instability. Moreover, LES is computationally expensive. Simpler frequency-based models are therefore often used in academia and industry for pre-design, optimization, control and uncertainty quantification. 

%\subsection{Thermo-acoustic frequency-based models}
%
%
There exist two main different classes of frequency-based low-order methods in thermo-acoustics.
\begin{enumerate}
\item 
Network-based methods model the geometry of the combustor as a network of  acoustic  elements where the acoustic problem can be solved analytically
\cite[]{Polifke2001,Stow2001,Dowling2003,Dowling2005}. Jump relations connect these elements, enforcing
 pressure continuity and mass or volume conservation~\cite{Bauerheim2015,Strobio2016} while accounting for the dilatation caused by flames. The acoustic quantities in each segment are related to the amplitudes of the forward and backward acoustic waves, which are determined such that all the jump relations and the boundary conditions are satisfied. This can only be achieved for discrete values of the complex eigenvalue.
A generic thermo-acoustic network is shown in Fig. \ref{fig:adjnet}a (adapted from Stow and Dowling~\cite{Stow2008}). 
\item 
Assuming a steady mean flow, which can be obtained by time-averaging experiments or numerical simulations, an equation for acoustic pressure perturbations in reacting flows can be derived from the Navier--Stokes equations \cite[]{Poinsot2005,Nicoud2007}. This is the (non-homogeneous) Helmholtz equation, where the source term stems from the heat released by the flame. Helmholtz solvers can calculate eigenvalues of complex three-dimensional configurations.
\begin{figure}
  \begin{center}  
 \includegraphics[scale=0.6]{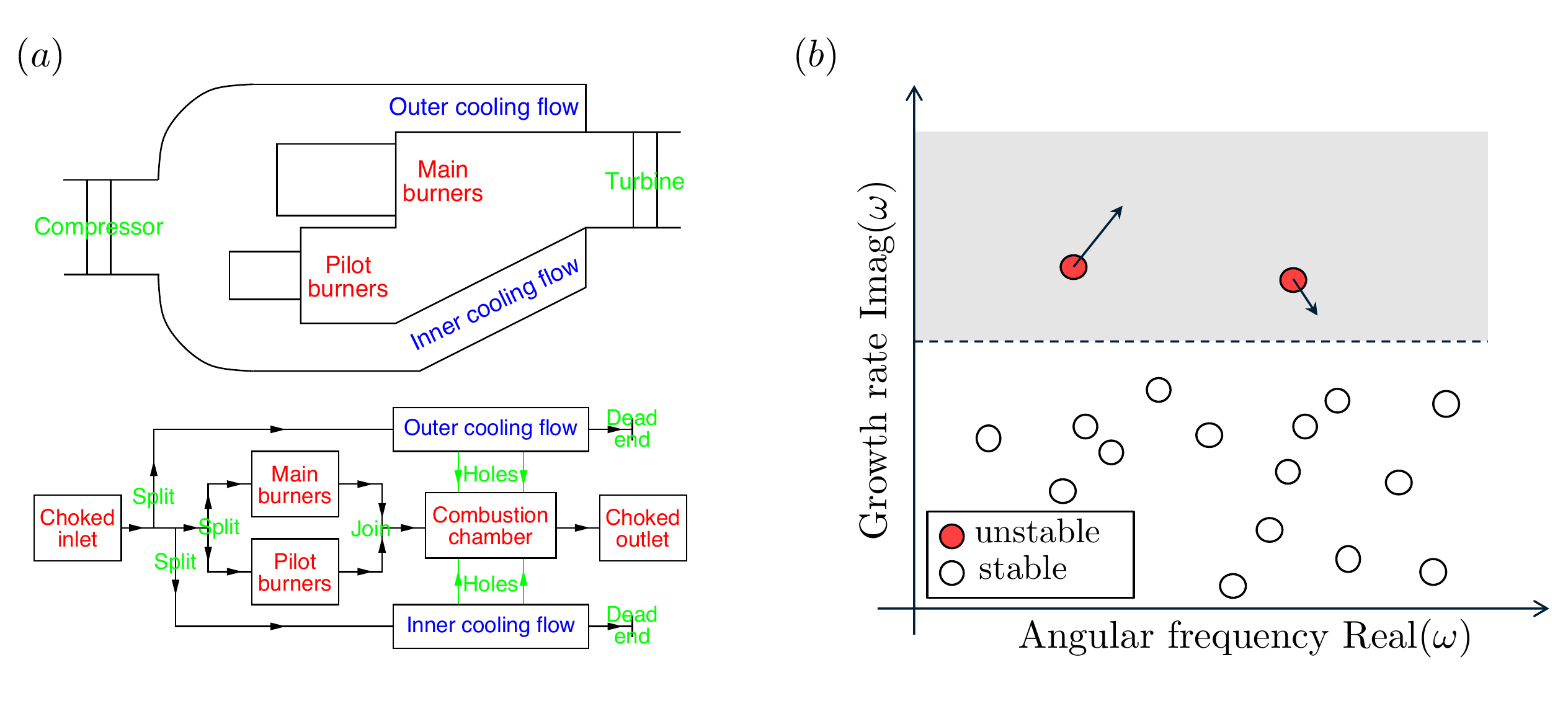}
  \end{center}
  \caption{(a) Generic thermo-acoustic network of an aero-engine. (Adapted from Stow and Dowling \cite{Stow2008} with permission of the original publisher ASME.) (b) If the system has some eigenvalues with positive growth rate (red/filled circles), the system is linearly  unstable.  Adjoint methods are able to quantify the effect that a change in any network parameter has on the eigenavalues of interest (arrows). This figure is for illustration and does not contain results from calculations.}\label{fig:adjnet}
  \end{figure}
\end{enumerate}
Using the modal transformation $\vecq(\mathbf{x},t)=\hat{\vecq}(\mathbf{x})\exp(-\mathrm{i}\omega t)$, both network-based models and Helmholtz approaches lead to an eigenvalue problem that is nonlinear in the eigenvalue
\begin{equation}
\label{eq:general}
\calN \left\{ \omega, \mathbf{p} \right\} \hat{\vecq} = 0, 
\end{equation}
where $\calN\{\}$ is an operator acting linearly on the thermo-acoustic eigenfunction, $\hat{\vecq}$, but depending nonlinearly on the complex eigenvalue, $\omega$. $\calN\{\}$ depends on the system's parameters\footnote{The thermo-acoustic parameters are also called `design' or `input' parameters in this paper.} (e.g., geometry, gain and phase of the flame response, area expansions), which are encapsulated in the vector $\mathbf{p}$. 
%The eigenvalue is an implicit function of the thermo-acoustic parameters $\mathbf{p}$ through the characteristic equation of the nonlinear eigenproblem \eqref{eq:general}, i.e., $\omega=\omega(\mathbf{p})$.  
The real part of the eigenvalue, $\omega_r$, is the angular frequency and the imaginary part, $\omega_i$, is the growth rate, which governs the stability. 
The size of the operator matrix is either equal to a few tens, in the case of a factorized network-based model, or proportional to the number of nodes in the numerical grid in the case of the Helmholtz approach, typically of order ten/hundred thousand for industrial geometries. If $\calN$ represents the Helmholtz problem, then the eigenfunction consists only of the discretized acoustic pressure. 

An important source of nonlinearity lies in the flame model, which introduces a time delay appearing as an exponential function in the frequency space \cite{Crocco1969}. Other nonlinearities in the eigenvalue may appear because of the boundary impedances \cite{Nicoud2007}. The solution of these nonlinear eigenproblems and the calculation of the thermo-acoustic growth rates and frequency is the objective of stability analysis. For design purposes, it is also important to predict how the thermo-acoustic stability changes due to  variations of the system. This is the objective of sensitivity analysis.
\subsection{Sensitivity analysis of eigenproblems}
%
%mode 
%
In situations that are susceptible to thermo-acoustic oscillations, often only a handful of oscillation modes are unstable. Existing techniques examine how a change in one parameter affects all oscillation modes, whether unstable or not.
 Adjoint techniques turn this around. In a single calculation, they examine how each oscillation mode is affected by changes in all parameters. In other words, they provide gradient information about the variation of an eigenvalue with respect to all the parameters in the model. 
For example, in a system with a thousand parameters, they calculate gradients a thousand times faster than finite-difference methods. 

Figure \ref{fig:adjnet}b is an illustration of the eigenvalues of a thermo-acoustic system. Two eigenmodes are unstable (they have positive growth rate and lie in the grey region). There are two approaches to determine how these two eigenvalues are affected by each system parameter. On the one hand, we could change each parameter independently and  recalculate all the eigenvalues, retaining only the information about the eigenvalues of interest. This is called the {\it finite-difference} approach in this paper and requires as many calculations as there are parameters. 
On the other hand, we could use adjoint methods to calculate how each eigenvalue is affected by every parameter, in a single calculation. This requires as many calculations as there are eigenvalues of interest, which is many times smaller than the number of parameters. 

Eigenvalue sensitivity methods originate from spectral perturbation theory \cite{Kato1980} and quantum mechanics \cite{Messiah1961}. In structural  mechanics, the calculation of first- and second-order derivatives of non-degenerate eigensolutions of self-adjoint nonlinear eigenproblems was proposed in aeroelasticity by Mantegazza and Bindolino~\cite{Mantegazza1987} and only theoretically by Liu and Chen~\cite{Liu1993}. Later, \cite{Jankovic1988,Andrew1990} found the analytical expressions for the sensitivities up to $n$-th order of a general self-adjoint non-degenerate eigenproblem with application to vibrational mechanics. More recently, Li et al. \cite{Li2013b} derived eigensolution sensitivities for self-adjoint problems with relevance both to degenerate and non-degenerate simplified structural mechanical problems. 
Eigenvalue sensitivity is also commonly used in hydrodynamic stability \cite{Tumin1984,Hill1992,Giannetti2007,Sipp2010a,Schmid2014,Tammisola2014a,Camarri2015}, where the eigenvalue problems are typically linear, or with quadratic nonlinearities, and non-degenerate. The review by Luchini and Bottaro \cite{Luchini2014} provides a thorough overview of the state-of-the-art of adjoint methods in hydrodynamic stability.

Adjoint eigenvalue sensitivity analysis of thermo-acoustic systems was proposed by Magri and Juniper \cite{Magri2013}. 
The analysis was applied to simplified models of combustors to find optimal passive mechanisms and sensitivity to base-state changes in a Rijke tube \cite{Magri2013,Magri2013e,MagriIJSCD,Rigas2015}, a ducted diffusion flame \cite{Magri2013c} and, more recently, to a ducted premixed-flame \cite{Orchini2015}. However, these studies dealt with linear eigenvalue problems in which the eigenvalue appears under a linear term. 

The extension of the adjoint analysis to nonlinear thermo-acoustic eigenproblems was proposed by Magri \cite{MagriPhD} and Juniper et al. \cite{Juniper2014ctr} based on ideas of spectral perturbation theory  of nonlinear eigenproblems \cite{Hinch1991}. 
They proposed two different adjoint methods for the prediction of eigenvalue sensitivities to perturbations to generic system's parameters. 
The first method was based on the Discrete Adjoint approach, in which the eigenvalue drift is obtained by recursive application of the linear adjoint formula at each iteration step of the nonlinear solver. The second approach was based on the linearization of the nonlinear operator around the unperturbed eigenvalue, which needs fewer operations than the first approach. In this paper we use the second approach of Juniper et al. \cite{Juniper2014ctr} and apply it to an elaborate annular combustor. Such first-order adjoint analysis was applied recently to predict symmetry breaking in annular combustors \cite{Mensah2015}. 

\subsection{Objective and Structure of the paper}
The aim of this paper is to provide a method for the calculation of first- and second-order eigenvalue sensitivities of non-self-adjoint nonlinear eigenproblems with degeneracy. This framework is applied to the elaborate annular combustor of \cite{Bauerheim2014a} to calculate design-parameter sensitivities. 

In Section \ref{sec:sens1} we present the theory for adjoint sensitivity analysis of nonlinear eigenproblems. We derive first- and second-order eigenvalue sensitivity relations both for non-degenerate and degenerate eigenproblems. 
%The key contribution is given by formulae \eqref{2eq:eigpert1} and \eqref{eq:degenerate}. 
In Section \ref{sec:appannu} the mathematical model of the annular combustor thermo-acoustic network  is briefly described.  For further background in annular combustors, the reader may refer to the review by O'Connor et al. \cite{OConnor2015}. 
In Section \ref{sec:results} we validate the adjoint formulae against finite differences, the latter of which provide the benchmark solution because they do not rely on any assumption on the perturbation size. Three configurations are considered: a weakly coupled rotationally symmetric combustor (Case A), a strongly coupled rotationally symmetric combustor (Case B)  and a strongly coupled non-rotationally symmetric combustor (Case C). 
The eigenvalue sensitivities to perturbations to both geometric, flow and flames parameters are calculated in Section \ref{sec:sens}. 
A concluding discussion ends the paper. 

All of these studies are based on deterministic analysis, which assumes perfect knowledge of the thermo-acoustic parameters. Including uncertainties in the flame parameters in the stability calculations is the objective of the second part of this paper \cite{Magri2016jcp2}. 
%%%%%%%%%%%%%%%%%%%%%%%%%%%%%%%
%%%%%%%%%%%%%%%%%%%%%%%%%%%%%%%
% THEORY
%%%%%%%%%%%%%%%%%%%%%%%%%%%%%%%
%%%%%%%%%%%%%%%%%%%%%%%%%%%%%%%
\section{Eigenvalue sensitivity of nonlinear eigenproblems}\label{sec:sens1}
We show how to compute the eigenvalue  sensitivity via equations involving the adjoint eigenfunctions.  
This approach combines a derivation with the Continuous Adjoint (CA) formulation, in which the problems are governed by continuous operators, without explicitly deriving the CA  equations. The final sensitivity equations can be applied by using a Discrete Adjoint (DA) philosophy, which is more accurate and easier to implement \cite[e.g., for thermo-acoustic problems,][]{Magri2013,MagriIJSCD,Magri2013c}. 

First, we solve for the nonlinear direct eigenproblem \eqref{eq:general}, in which the eigenvalue appears under exponential, polynomial and rational terms. Starting from an initial guess for the eigenvalue, we assume that the converged eigenvalue $\omega_0$ is a numerical root of the dispersion relation 
\begin{align}\label{eq:det}
&  \lvert\textrm{det}\left(\calN\left\{\omega_0,\mathbf{p}_0\right\}\right)\lvert<\textrm{tol},
\end{align}
where `$\textrm{det}$' is the determinant and `tol' is a desired tolerance. In large systems we ensure condition \eqref{eq:det} through relaxation methods \cite{Nicoud2007} instead of solving for the characteristic equation. 
Equation \eqref{eq:det} defines an implicit function between $\omega$ and $\mathbf{p}$, i.e., $\omega=\omega(\mathbf{p})$. 
The corresponding eigenfunction $\hat{\vecq}_0$ is calculated from the linear system
% \lm{MB: I don't know if you see my comments the last time. I think we should be very careful with the notations, especially making a distinction between q0 q1 and q etc. For what I understand, when the operator is N(omega0,p0), the eigenvector has to be q0, not q}
\begin{align}\label{eq:eig1}
&  \calN\left\{\omega_0,\mathbf{p}_0\right\}\hat{\vecq}_0=0.
\end{align}
The operator $\calN$ depends only on the final converged eigenvalue, $\omega_0$. 
%The linear system $\eqref{eq:eig1}$ has infinite solutions because the operator is singular. 
%\textcolorblack{[Is this correct? The null space of $\left(\calN\left\{\omega_N,\mathbf{p}\right\}-\mathcal{B}\right)$ should be a unique vector (the eigenvector). Isn't therefore only one non-trivial solution?]}
The kernel of equation \eqref{eq:eig1} can be found by computing the  singular vector(s) associated with the trivial singular value(s). 

Second, by defining the adjoint eigenfunction, $\hat{\vecq}_0^+$, and operator through a Hermitian inner product in an appropriate Hilbert space
\begin{align}\label{eq:innpr}
& \left\langle\hat{\vecq}_0^+, \calN\left\{\omega_0,\mathbf{p}_0\right\}\hat{\vecq}_0\right\rangle=\left\langle\calN\left\{\omega_0,\mathbf{p}_0\right\}^+\hat{\vecq}_0^+, \hat{\vecq}_0\right\rangle,
\end{align}
we solve for the adjoint eigenfunction associated with the converged eigenvalue $\omega_0$
\begin{align}\label{eq:adeig}
& \calN\left\{\omega_0,\mathbf{p}_0\right\}^H\hat{\vecq}_0^{+}=0.  
\end{align}
%CHECK IF THERE IS SOME COMPLEX CONJUGATION MISSING 
%

%The complex conjugation is denoted by *. 
If we followed a purely Continuous Adjoint (CA) approach \citep{Magri2013,MagriIJSCD,Magri2013e}, we would need to derive explicitly the Hermitian operator $\calN^H$ and the continuous adjoint equations. However, we do not derive these equations explicitly and we proceed on only with the abstract expression of the Hermitian operator, in order to apply the Discrete Adjoint (DA) method directly to the final sensitivity relations, as explained subsequently. 
In equation \eqref{eq:adeig}, the adjoint eigenfunction can be found with the same procedure as \eqref{eq:eig1}. 
%
%Also, we assume that the eigenproblem is non degenerate, meaning that the eigenvalues' geometric multiplicity is unity. The extension to degenerate eigenproblem is straightforward \citep{Hinch1991} and takes advantage of the Fredholm alternative presented afterwards.
% 
Third, we perturb a system's parameter and calculate the perturbation operator, which we can evaluate numerically by finite difference 
\begin{align}\label{2eq:pert}
& \mathbf{p}=\mathbf{p}_0+\epsilon\mathbf{p}_1\implies \delta_p\calN\{\omega_0,\epsilon\mathbf{p}_1\}=  \calN\{\omega_0,\mathbf{p}\} -\calN\{\omega_0,\mathbf{p}_0\}, 
\end{align}
where  $\epsilon\ll 1$. This perturbation operator is the input of the problem and, therefore, is constant, i.e., it does not depend on $\omega$. 
Hence, $\delta_p\calN\{\omega_0, \epsilon\mathbf{p}_1\}$ represents exactly all the orders of its Taylor series (providing that $\epsilon\mathbf{p}_1$ is sufficiently small)
\begin{align}\label{eq:100}
\delta_p\calN\{\omega_0,\epsilon\mathbf{p}_1\}=  \frac{\partial\calN}{\partial\mathbf{p}}\epsilon\mathbf{p}_1+\frac{1}{2}\frac{\partial^2\calN}{\partial\mathbf{p}^2}(\epsilon\mathbf{p}_1)^2+o(\epsilon^2). 
\end{align}
We assume that the eigenvalues and eigenfunctions are analytical in the complex plane around $\epsilon=0$ and
\begin{align}\label{2eq:pert2}
&\omega=\omega_0+(\epsilon\mathbf{p}_1)\omega_1+\frac{1}{2}(\epsilon\mathbf{p}_1)^2\omega_2 \textrm{\;\;\;\;\;\;and\;\;\;\;\;\;}\hat{\vecq}=\hat{\vecq}_0+(\epsilon\mathbf{p}_1)\hat{\vecq}_1+\frac{1}{2}(\epsilon\mathbf{p}_1)^2\hat{\vecq}_2,
\end{align}
where 
\begin{align}\label{2eq:pert200}
 \omega_1 = \frac{\mathrm{d}\omega}{\mathrm{d}\mathbf{p}},\;\; \omega_2 = \frac{\mathrm{d}^2\omega}{\mathrm{d}\mathbf{p}^2} \textrm{\;\;\;\;\;\;and\;\;\;\;\;\;}\hat{\vecq}_1 = \frac{\mathrm{d}\hat{\vecq}}{\mathrm{d}\mathbf{p}}, \;\; \hat{\vecq}_2 = \frac{\mathrm{d}^2\hat{\vecq}}{\mathrm{d}\mathbf{p}^2}. 
\end{align}
%
%The perturbation operator $\delta_p\calN$ is calculated by finite difference from \eqref{2eq:pert}. 
%
%
The perturbed eigenproblem must satisfy equation \eqref{eq:general} and is Taylor-expanded up to the second-order total derivative of $\mathbf{p}$ around the unperturbed eigenvalue $\omega_0$, yielding 
\begin{align}
& \calN\left\{\omega_0+(\epsilon\mathbf{p}_1)\omega_1+\frac{1}{2}(\epsilon\mathbf{p}_1)^2\omega_2, \mathbf{p}_0+\epsilon\mathbf{p}_1\right\}\left(\hat{\vecq}_0+(\epsilon\mathbf{p}_1)\hat{\vecq}_1+\frac{1}{2}(\epsilon\mathbf{p}_1)^2\hat{\vecq}_2\right)=0, \nonumber \\
&\implies \calN\left\{\omega_0, \mathbf{p}_0\right\}\hat{\vecq}_0 + \frac{\mathrm{d}\calN\left\{\omega, \mathbf{p}\right\}\hat{\vecq}}{\mathrm{d}\mathbf{p}}(\epsilon\mathbf{p}_1)+ \frac{1}{2}\frac{\mathrm{d}^2\calN\left\{\omega, \mathbf{p}\right\}\hat{\vecq}}{\mathrm{d}\mathbf{p}^2}(\epsilon\mathbf{p}_1)^2 + o(\epsilon^2)=0. \label{2eq:eigpert22}
\end{align}
By taking the total derivatives and using definitions \eqref{eq:100} and \eqref{2eq:pert200}, we obtain the order-by-order expansion 
\begin{align}
&  \calN\left\{\omega_0, \mathbf{p}_0\right\}\hat{\vecq}_0  + \nonumber \\ 
&+ (\epsilon\mathbf{p}_1)\left[\calN\left\{\omega_0, \mathbf{p}_0\right\}\hat{\vecq}_1 + \frac{\partial\calN\left\{\omega, \mathbf{p}_0\right\}}{\partial\omega}\Big\lvert_{\omega_0}\omega_1\hat{\vecq}_0+ \delta_p\calN\left\{\omega_0, \epsilon\mathbf{p}_1\right\}\hat{\vecq}_0 \right] +\nonumber\\
& +(\epsilon\mathbf{p}_1)^2\left[\frac{1}{2}\calN\left\{\omega_0, \mathbf{p}_0\right\}\hat{\vecq}_2 +\frac{\partial\calN\left\{\omega, \mathbf{p}_0\right\}}{\partial\omega}\Big\lvert_{\omega_0}\omega_1\hat{\vecq}_1+ \delta_p\calN\left\{\omega_0, \epsilon\mathbf{p}_1\right\}\hat{\vecq}_1\right]+\nonumber\\
&+(\epsilon\mathbf{p}_1)^2\left[ \frac{1}{2}\frac{\partial^2\calN\left\{\omega,\mathbf{p}_0\right\}}{\partial\omega^2}\Big\lvert_{\omega_0}\omega_1^2 +  \frac{1}{2}\frac{\partial\calN\left\{\omega,\mathbf{p}_0\right\}}{\partial\omega}\Big\lvert_{\omega_0}\omega_2  + \frac{\partial\delta_p\calN\left\{\omega_0,\epsilon\mathbf{p}_1\right\}}{\partial\omega}\Big\lvert_{\omega_0}\omega_1\right]\hat{\vecq}_0+o(\epsilon^2)= 0. \label{2eq:eigpert1}
\end{align}
Importantly, the cross derivative $\partial\delta_p\calN\left\{\omega_0,\epsilon\mathbf{p}_1\right\}/\partial\omega$ is zero because the perturbation operator $\delta_p\calN\left\{\omega_0,\epsilon\mathbf{p}_1\right\}$ is constant.  
%\textcolor{red}{Note that the red term is zero as $p_2$ is zero. I kept it for fun.} 
The unperturbed term $\sim\mathcal{O}(1)$ in equation \eqref{2eq:eigpert1} is trivially zero because of equation  \eqref{eq:eig1}. 
Higher order terms $\sim o(\epsilon^2)$ are neglected. 
%\footnote{Here, the `big-O' notation represents a quantity of the same order as its argument.}

\subsection{First-order eigenvalue sensitivity}\label{sec:omega1}
The equation for the first order $\sim\mathcal{O}(\epsilon)$ is recast as
\begin{align}\label{2eq:eigpert3}
& \calN\left\{\omega_0, \mathbf{p}_0\right\}\hat{\vecq}_1 = - \left( \frac{\partial\calN\left\{\omega, \mathbf{p}_0\right\}}{\partial\omega}\Big\lvert_{\omega_0}\omega_1\hat{\vecq}_0+ \delta_p\calN\left\{\omega_0, \epsilon\mathbf{p}_1\right\}\hat{\vecq}_0\right).
\end{align}
The objective is to find the eigenvalue drift $\omega_1$  due to the perturbation $\delta_p\calN$. 
The adjoint eigenfunction provides a solvability condition for the non-homogeneous system \eqref{2eq:eigpert3} fulfilling the Fredholm alternative\footnote{The left-hand side operator range is equal to the kernel of the orthogonal complement of its adjoint operator. } \cite{Oden1979}.  Mathematically, this is achieved by projecting equation \eqref{2eq:eigpert3} onto the adjoint eigenfunction, $\hat{\vecq}_0^+$   
\begin{align}\label{2eq:eigpert4}
& \left\langle\hat{\vecq}_0^+, \calN\left\{\omega_0, \mathbf{p}_0\right\}\hat{\vecq}_1\right\rangle = 
-  \left\langle\hat{\vecq}_0^+,\left( \frac{\partial\calN\left\{\omega, \mathbf{p}_0\right\}}{\partial\omega}\Big\lvert_{\omega_0}\omega_1\hat{\vecq}_0+ \delta_p\calN\left\{\omega_0, \epsilon\mathbf{p}_1\right\}\hat{\vecq}_0\right)\right\rangle.
\end{align}
Using equation \eqref{eq:adeig}, the definition of the inner product \eqref{eq:innpr} and its linearity, yields an equation for the first-order eigenvalue drift 
\begin{align}\label{2eq:eigpert5}
& \omega_1 = 
 \frac{- \left\langle\hat{\vecq}_0^+, \delta_p\calN\left\{\omega_0, \epsilon\mathbf{p}_1\right\}\hat{\vecq}_0\right\rangle}{\left\langle\hat{\vecq}_0^+, \frac{\partial\calN\left\{\omega, \mathbf{p}_0\right\}}{\partial\omega}\Big\lvert_{\omega_0}\hat{\vecq}_0\right\rangle}, 
\end{align}
assuming that $\partial\calN\left\{\omega,\mathbf{p}_0\right\}/\partial\omega\not=0$. 
% the denominator is different from zero because of the bi-orthogonality condition \cite{Salwen1981,MagriIJSCD}.
If  the number of components of $\mathbf{p}$ is $S$, and we are interested in the first-order sensitivity for each,  equation \eqref{2eq:eigpert5} enables us to reduce the number of nonlinear-eigenproblem computations by circa $SP$, where $P$ is the average of the number of iterations needed to obtain $\omega_1$ by solving the nonlinear eigenproblem perturbed via finite difference. \\

If the unperturbed eigenvalue $\omega_0$ is $N$-fold degenerate\footnote{$N$-fold degeneracy occurs when an eigenvalue has $N$ independent associated eigenfunctions, i.e., the eigenvalue has $N$ geometric multiplicity.}, the eigenfunction expansion becomes $\hat{\vecq}=\sum_{i=1}^N\alpha_i\hat{\mathbf{e}}_{0,i} + \epsilon\hat{\vecq}_1+ \epsilon^2\hat{\vecq}_2$, where $\alpha_i$ are complex numbers and $\hat{\mathbf{e}}_{0,i}$ are the $N$ independent eigenfunctions associated with $\omega_0$. By requiring the right-hand side of equation \eqref{2eq:eigpert4} to have no components along the independent directions $\hat{\mathbf{e}}_{0,i}$ (Fredholm alternative), we obtain an eigenproblem in $\alpha_i$ and eigenvalue $\omega_1$ \citep{Hinch1991}
\begin{align}\label{eq:degenerate}
& \left\langle\hat{\mathbf{e}}^+_{0,i},\frac{\partial\calN\left\{\omega, \mathbf{p}_0\right\}}{\partial\omega}\Big\lvert_{\omega_0}\hat{\mathbf{e}}_{0,j}\right\rangle\omega_1\alpha_j = -\left\langle\hat{\mathbf{e}}^+_{0,i}, \delta_p\calN\left\{\omega_0, \epsilon\mathbf{p}_1\right\}\hat{\mathbf{e}}_{0,j}\right\rangle\alpha_j,
\end{align}
for $i,j=1,2,...,N$. Einstein summation is used, therefore, the inner products in equation \eqref{eq:degenerate} are the components of an $N\times N$ matrix, $\alpha_j$ are the components of an $N\times 1$ vector and $\omega_1$ is the eigenvalue. This equation is defined only in the $N$-fold degenerate subspace. 
In thermo-acoustics, degeneracy occurs in rotationally symmetric annular combustors in which azimuthal modes have 2-fold degeneracy \cite{Noiray2013a,Ghirardo2013a,Bauer2014_jfm,Mensah2015}. 
The generalized eigenproblem \eqref{eq:degenerate} outputs $N$ first-order eigenvalue drifts and $N$ unperturbed eigendirections. 
%We are interested in calculating how the stability of the system is affected by a perturbation in the operator, therefore, 
We select the first-order eigenvalue drift, $\omega_1$, with greatest growth rate, which causes the greatest change in the stability. \\

To demonstrate the adjoint-based eigenvalue sensitivity  \eqref{2eq:eigpert5}, we consider the generic nonlinear eigenvalue problem represented by a $2\times2$ matrix
\begin{align}\label{eq:ped:dir}
\left( \begin{matrix} % or pmatrix or bmatrix or Bmatrix or ...
      \calN_{11}(\omega) &\calN_{12}(\omega)\\
      \calN_{21}(\omega) & \calN_{22}(\omega)\\
   \end{matrix}\right) 
   \left( \begin{matrix} % or pmatrix or bmatrix or Bmatrix or ...
      \qh_1 \\
      \qh_2\\
   \end{matrix}\right) 
   = 
      \left( \begin{matrix} % or pmatrix or bmatrix or Bmatrix or ...
       0 \\
       0 
   \end{matrix}\right).
\end{align}
We solve for the characteristic equation 
\begin{align}\label{eq:ped:char}
\textrm{F}(\omega) = \calN_{11}(\omega)\calN_{22}(\omega) - \calN_{21}(\omega)\calN_{12}(\omega)=0,
\end{align}
and find $\omega_0$ such that $\mathrm{F}(\omega_0)=0$. We assume that this root is non degenerate. 
The corresponding direct and adjoint eigenvectors are, respectively 
\begin{align}\label{eq:ped:eigenvecs}
\hat{\vecq}_0 = 
   \left( \begin{matrix} % or pmatrix or bmatrix or Bmatrix or ...
      -\calN_{12}(\omega_0)/\calN_{11}(\omega_0) \\
      1 \\
   \end{matrix}\right) \qh_2, \\
   \hat{\vecq}^+_0 = 
   \left( \begin{matrix} % or pmatrix or bmatrix or Bmatrix or ...
      -\calN_{21}(\omega_0)^*/\calN_{11}(\omega_0)^* \\
      1 \\
   \end{matrix}\right) \qh^+_2, 
\end{align}
where $\calN_{11}(\omega_0)$ is assumed $\not=0$ and $\qh_2$, $\qh^+_2$ are arbitrary non-trivial complex numbers, which are set to 1. 
The dependency on $\omega_0$ is dropped for brevity from now on. 
Assuming that the characteristic equation defines a continuously differentiable manifold, the exact first-order eigenvalue sensitivity is calculated by the implicit function theorem (also known as Dini's theorem)
\begin{align}\label{eq:ped:dini}
\frac{\partial\omega}{\partial p} &= -\frac{\partial F/\partial p }{\partial F/\partial \omega}\nonumber \\
&=-\frac{\calN_{11}\partial\calN_{22}/\partial p + \calN_{22}\partial\calN_{11}/\partial p - \calN_{12}\partial\calN_{21}/\partial p - \calN_{21}\partial\calN_{12}/\partial p}{\calN_{11}\partial\calN_{22}/\partial\omega + \calN_{22}\partial\calN_{11}/\partial\omega - \calN_{12}\partial\calN_{21}/\partial\omega - \calN_{21}\partial\calN_{12}/\partial\omega}. 
\end{align}
Using an Euclidean Hermitian inner product, the adjoint eigenvalue sensitivity \eqref{2eq:eigpert5}, for this algebraic problem, reads
\begin{align}\label{eq:ped:adj}
\frac{\delta\omega}{\delta p} &= - \frac{\hat{\vecq}_0^{+H} (\partial\calN/\partial p) \hat{\vecq}_0}{\hat{\vecq}_0^{+H} (\partial\calN/\partial\omega) \hat{\vecq}_0}, \nonumber \\
& = - \frac{ 
   \left( \begin{matrix} % or pmatrix or bmatrix or Bmatrix or ...
      -\calN_{21}^*/\calN_{11}^* \;\;\;\;
      1
   \end{matrix}\right)^*   
   \left( \begin{matrix} % or pmatrix or bmatrix or Bmatrix or ...
      \partial\calN_{11}/\partial p & \partial\calN_{12}/\partial p \\
       \partial\calN_{21}/\partial p & \partial\calN_{22}/\partial p 
   \end{matrix}\right)
   \left( \begin{matrix} % or pmatrix or bmatrix or Bmatrix or ...
      -\calN_{12}/\calN_{11} \\
      1
   \end{matrix}\right)}{   
   \left( \begin{matrix} % or pmatrix or bmatrix or Bmatrix or ...
      -\calN_{21}^*/\calN_{11}^*\;\;\;\;
      1
   \end{matrix}\right)^*  
      \left( \begin{matrix} % or pmatrix or bmatrix or Bmatrix or ...
      \partial\calN_{11}/\partial\omega& \partial\calN_{12}/\partial\omega \\
       \partial\calN_{21}/\partial\omega & \partial\calN_{22}/\partial\omega 
   \end{matrix}\right)
   \left( \begin{matrix} % or pmatrix or bmatrix or Bmatrix or ...
      -\calN_{12}/\calN_{11}\\
      1
   \end{matrix}\right)}. 
\end{align}
 When the vector-matrix-vector multiplications are performed, the adjoint-based sensitivity \eqref{eq:ped:adj} coincides with the analytical sensitivity \eqref{eq:ped:dini}. This illustrates that equation (\ref{2eq:eigpert5}) is an exact representation of the first-order eigenvalue drift, $\delta \omega / \delta p$.

\subsection{Second-order eigenvalue sensitivity}\label{sec:appendix}
The equation for the second-order is recast as 
\begin{align}\nonumber
& \frac{1}{2}\calN\left\{\omega_0, \mathbf{p}_0\right\}\hat{\vecq}_2= -\left(\frac{\partial\calN\left\{\omega, \mathbf{p}_0\right\}}{\partial\omega}\Big\lvert_{\omega_0}\omega_1\hat{\vecq}_1+ \delta_p\calN\left\{\omega_0, \epsilon\mathbf{p}_1\right\}\hat{\vecq}_1\right)+\\
&-\left[ \frac{1}{2}\frac{\partial^2\calN\left\{\omega,\mathbf{p}_0\right\}}{\partial\omega^2}\Big\lvert_{\omega_0}\omega_1^2 +  \frac{1}{2}\frac{\partial\calN\left\{\omega,\mathbf{p}_0\right\}}{\partial\omega}\Big\lvert_{\omega_0}\omega_2  \right]\hat{\vecq}_0=0.\label{2eq:eigpert2}
\end{align}
The calculation of the second-order eigenvalue drift is obtained by projecting equation \eqref{2eq:eigpert2} onto the adjoint eigenfunction, yielding
\begin{align}\nonumber
&\frac{1}{2} \left\langle\hat{\vecq}^+_0, \calN\left\{\omega_0, \mathbf{p}_0\right\}\hat{\vecq}_2\right\rangle = 
  \left\langle\hat{\vecq}^+_0,-\left(\frac{\partial\calN\left\{\omega, \mathbf{p}_0\right\}}{\partial\omega}\Big\lvert_{\omega_0}\omega_1\hat{\vecq}_1+ \delta_p\calN\left\{\omega_0, \epsilon\mathbf{p}_1\right\}\hat{\vecq}_1\right)\right\rangle+\\
&\left\langle\hat{\vecq}^+_0,-\left[ \frac{1}{2}\frac{\partial^2\calN\left\{\omega,\mathbf{p}_0\right\}}{\partial\omega^2}\Big\lvert_{\omega_0}\omega_1^2 +  \frac{1}{2}\frac{\partial\calN\left\{\omega,\mathbf{p}_0\right\}}{\partial\omega}\Big\lvert_{\omega_0}\omega_2  \right]\hat{\vecq}_0\right\rangle.\label{2eq:eigpert7}
\end{align}
Using equations \eqref{eq:adeig} and \eqref{eq:innpr} yields an equation for the second-order eigenvalue drift 
\begin{align}\nonumber
& \omega_2 = 
-  2\frac{\left\langle\hat{\vecq}^+_0,\left(\frac{\partial\calN\left\{\omega, \mathbf{p}_0\right\}}{\partial\omega}\Big\lvert_{\omega_0}\omega_1\hat{\vecq}_1+ \delta_p\calN\left\{\omega_0, \epsilon\mathbf{p}_1\right\}\hat{\vecq}_1\right)\right\rangle}{\left\langle\hat{\vecq}^+_0,\frac{\partial\calN\left\{\omega,\mathbf{p}_0\right\}}{\partial\omega}\Big\lvert_{\omega_0}\hat{\vecq}_0\right\rangle} +  \\
& -2\frac{\left\langle \hat{\vecq}^+_0,\left(\frac{1}{2}\frac{\partial^2\calN\left\{\omega,\mathbf{p}_0\right\}}{\partial\omega^2}\Big\lvert_{\omega_0}\omega_1^2  \right)\hat{\vecq}_0\right\rangle}{\left\langle\hat{\vecq}^+_0,\frac{\partial\calN\left\{\omega,\mathbf{p}_0\right\}}{\partial\omega}\Big\lvert_{\omega_0}\hat{\vecq}_0\right\rangle}.\label{2eq:eigpert8}
\end{align}
The eigenvalue-drift equations \eqref{2eq:eigpert5},\eqref{eq:degenerate},\eqref{2eq:eigpert8} enable the calculation of the $i$-th drift only by using eigenfunctions up to $(i-1)$-th order. The calculation of the perturbed eigenfunction $\hat{\vecq}_1$, necessary for the calculation of the second-order eigenvalue drift, is discussed in the next section. 
%^
%
\subsection{Calculation of the perturbed eigenfunction}
\label{sec:PerEigf}

The calculation of the perturbed eigenfunction $\hat{\vecq}_1$ in equation  \eqref{2eq:eigpert3} requires solving for a non-homogeneous singular linear system because the inversion operator, $\calN^{-1}\{\omega_0,\mathbf{p}_0\}$, does not exist. However, the compatibility condition ensures that this linear system has (infinite) solutions. %  $\left\langle\hat{\vecq}^+, \mathbf{\Psi}\right\rangle=0$ 
For brevity, we define $\dim(\calN)=K$ and use matrices. 
In a non-defective degenerate system, a complete eigenbasis is $\{\hat{\vecq}_0,\hat{\mathbf{e}}_{i}\}$, where $i=1,2,\ldots,K-N$ and $\hat{\vecq}_0=\sum_{j}^N\alpha_j\hat{\mathbf{e}}_{0,j} $ and $\hat{\mathbf{e}}_{i}$ are the remaining non-degenerate eigenfunctions. (We are assuming that only the $0$-th eigenfunction is $N$-fold degenerate. The extension to other eigenfunctions' degeneracy is straightforward.) In general, the coefficients $\alpha_j$ are arbitrary, however, when working with perturbations, these coefficients are uniquely determined by the first-order sensitivity \eqref{eq:degenerate}. 
The perturbed eigenfunction is decomposed as 
\begin{align}\label{eq:nelsdec}
\hat{\vecq}_1 = \hat{\mathbf{z}} + \beta_0\hat{\vecq}_0, 
\end{align}
where $\beta_0$ is in general a complex number. 
By substituting equation  \eqref{eq:nelsdec} into \eqref{2eq:eigpert3}, we obtain 
\begin{align}\label{eq:q1dec}
\calN\{\omega_0,\mathbf{p}_0\}\hat{\mathbf{z}} = \mathbf{\Psi}, 
\end{align}
because $\calN\{\omega_0,\mathbf{p}_0\}(\hat{\mathbf{z}}+ \beta_0\hat{\vecq}_0)=\calN\{\omega_0,\mathbf{p}_0\}\hat{\mathbf{z}}$ for equation \eqref{eq:eig1}. 
$\mathbf{\Psi}$ is the right-hand side of equation \eqref{2eq:eigpert3}. 
$\hat{\vecq}_1$ is then calculated as follows. 
\begin{itemize}
\item Decompose  $\calN=\mathbf{U}\tilde{\mathbf{\Lambda}}\mathbf{V}^{H}$ (Singular Value Decomposition, SVD), where 
\begin{align}
\tilde{\mathbf{\Lambda}} = \left( \begin{matrix} 
      \mathbf{\Lambda} & \mathbf{0}\\
      \mathbf{0} & \mathbf{\Theta}
   \end{matrix}\right). 
\end{align}
The submatrix $\mathbf{\Lambda}$ is diagonal and contains the $K-N$ non-trivial singular values of $\calN\{\omega_0, \mathbf{p}_0\}$. The submatrix $\mathbf{\Theta}$ is a $N\times N$ null matrix. 
The columns of the unitary matrix $\mathbf{U}$ are the left singular vectors and the columns of the unitary matrix $\mathbf{V}$ are the right singular vectors. 
\item 
Set $\mathbf{\Theta}$ to any non-trivial diagonal matrix, for example, the identity matrix. 
\item 
Solve for 
\begin{align}
\left( \begin{matrix} 
      \mathbf{Y}_1 \\
      \mathbf{Y}_2
   \end{matrix}\right) = \tilde{\mathbf{\Lambda}}^{-1}\mathbf{U}^{-1}\mathbf{\Psi}. 
\end{align}
\item 
Set $\mathbf{Y}_2=0$ and find the solution 
\begin{align}
\hat{\mathbf{z}}=\mathbf{V}\left( \begin{matrix} 
      \mathbf{Y}_1 \\
      0
   \end{matrix}\right).
\end{align}
\end{itemize}

Another method for the calculation of $\hat{\vecq}_1$ is presented in Appendix~A.\\
%
%
%
%

%
%\subsubsection{The coefficient $\beta_0$}\label{sec:Beta0}
%The  coefficient $\beta_0$ is determined by normalizing 
%However, if a normalization has to be imposed, $\beta_0$ can be determined uniquely. 
%The normalization is assumed to be $\hat{\vecq}_0^H\mathbf{M}\hat{\vecq}_0=b$, where $\mathbf{M}$ is a symmetric matrix and $b\in\mathbb{R}$. 
%The first-order perturbed normalization gives $Re(\hat{\vecq}_0^H\mathbf{M}\hat{\vecq}_1)=0$. 
%By substituting equation \eqref{eq:nelsdec} into the previous equation, we obtain the coefficient 
%\begin{align}
%\beta_0 = -\frac{Re\left(\hat{\vecq}^H\mathbf{M}\hat{\mathbf{z}}\right)}{b}. 
%\end{align}
%
In this study no normalization constraint is imposed and, therefore, $\beta_0$ is arbitrarily set to zero. This means that we are removing the non-uniqueness of $\hat{\vecq}_1$ by requiring it not to have a component along the unperturbed eigenfunction $\hat{\vecq}_0$ \cite{Hinch1991}. 
%%%%%%%%%%%%%%%%%%%%%%%%%%%%%%%
%%%%%%%%%%%%%%%%%%%%%%%%%%%%%%%
% ANNULAR COMBUSTOR 
%%%%%%%%%%%%%%%%%%%%%%%%%%%%%%%
%%%%%%%%%%%%%%%%%%%%%%%%%%%%%%%
%!TEX root = ./MAIN.tex

\section{Mathematical model of an annular combustor}\label{sec:appannu}
\begin{figure}[!htb]
  \begin{center} 
 \includegraphics[scale=0.4]{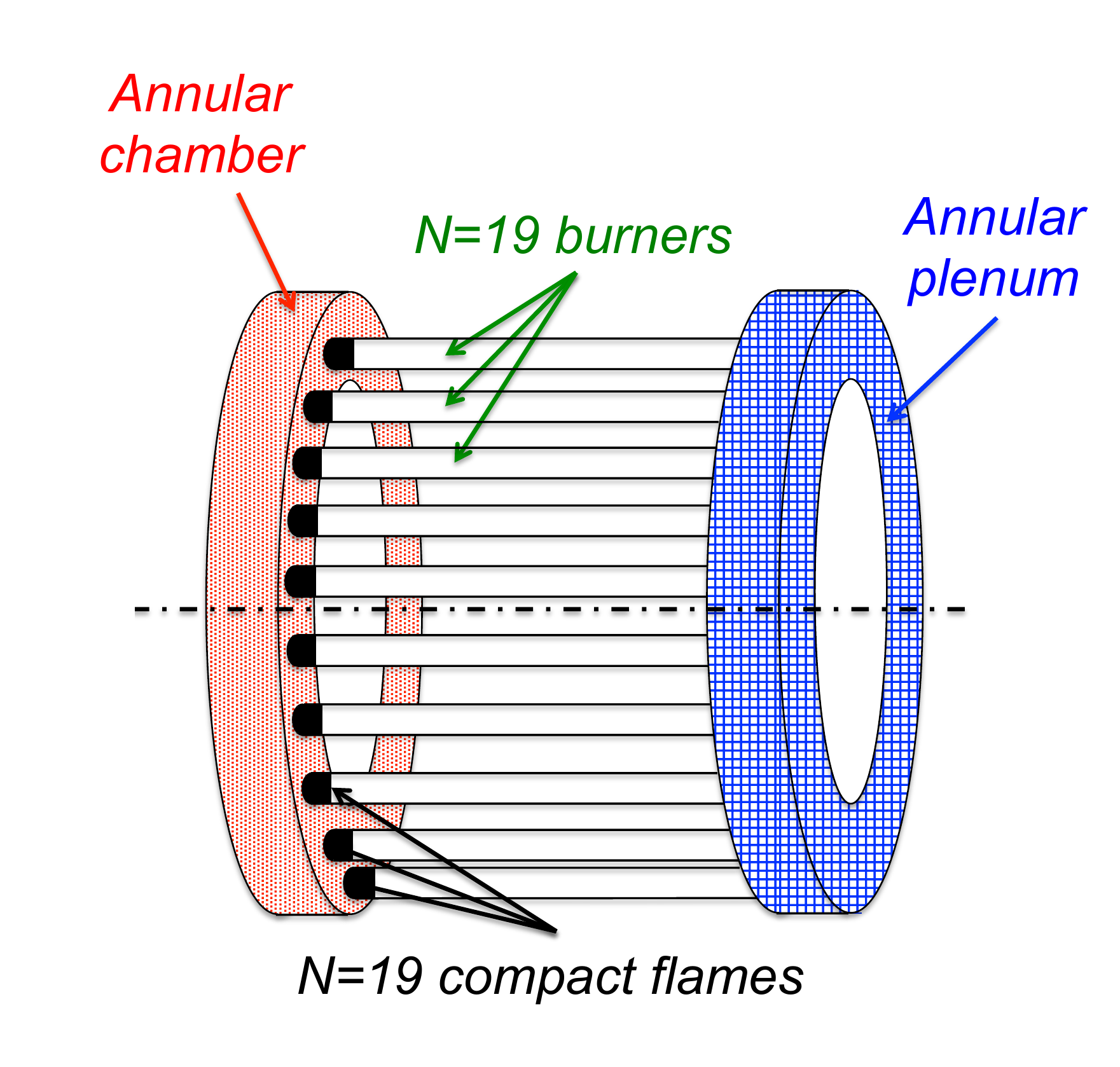}
  \end{center}
  \caption{Schematic of the annular combustor, which consists of a plenum and combustion chamber connected by longitudinal burners \citep{Bauerheim2014a,Bauerheim2016}. }\label{fig:network}
\end{figure}

%
%\lm{MB: I change the title to make a clear difference with the next section dealing with UQ.}
Annular combustion chambers are commonly used in aircraft gas turbines because of their compactness and ability for efficient light around~\cite{Boileau2008,OConnor2015}. Such configurations, however, suffer from combustion instabilities due to azimuthal modes that often appear at low frequencies, where damping mechanisms are less effective. We study  an annular combustor configuration typical of modern ultra Low-NOx combustion chambers, detailed in~\cite{Bauerheim2016}. The network model developed by Bauerheim et al. \cite{Bauerheim2014a}, which was validated against a three-dimensional Helmholtz solver to predict the stability of azimuthal modes, is therefore used in the present study. This low-order model describes a combustion chamber connected by longitudinal burners fed by a common annular plenum (Fig. \ref{fig:network}). 
%
%\subsection{The mathematical model}\label{sec:NetworkModel}
%

%
%We look for eigenpairs $\left(\hat{\vecq},\omega\right)\in\mathbb{C}^4\times\mathbb{C}$ such that 
The Annular Network Reduction methodology \cite{Bauerheim2014a} analytically derives the dispersion relation $\det \left( \calN\{\omega,\mathbf{p}\} \right)$ $=$ $0$ of the annular system, where the operator $\calN$ is defined as 
\begin{align}
\calN\{\omega,\mathbf{p}\}=\prod_i^{N_b}\calR_i(\omega)\calT_i\left(\omega,\mathbf{p}\right)-\calI,\label{eq:mod1}
\end{align}
where $\mathcal{I}$ is the identity operator and $N_b$=19 is the number of burners. $\calR_i\in\mathbb{R}^{4\times4}$ is the propagation operator that maps the acoustic waves in the uniform components of the network, represented by the matrices 
\begin{align}
 \calR_i&=\left( \begin{matrix} 
      \matR(k_p\Delta x_p) & \mathbf{0}\\
      \mathbf{0} & \matR(k_c\Delta x_c)\\
   \end{matrix}\right), \\
\matR(k\Delta x)&=\left( \begin{matrix} 
      \cos(k\Delta x) & -\sin(k\Delta x)\\
      \sin(k\Delta x) & \cos(k\Delta x)\\
   \end{matrix}\right), 
\end{align}
where $k$ is the wavenumber and $\Delta x$ is the annular abscissa. The subscripts $p$ and $c$ stand for `plenum' and `combustion chamber', respectively. The vector of the thermo-acoustic parameters is $\mathbf{p}$ $=$ $(\{n_i,$ $\tau_i\},$ $\alpha,$ $L_p,$ $S_p,$ $L_i,$ $S_i,$ $\rho_p,$ $\rho_c$, $c_c$, $c_p$), where $i=1,2,\ldots,N$; $n_i$ are the flame gains (or indices) and $\tau_i$ are the time delays; $\alpha$ is the flame position within the burner, which is the same for all the flames; $L_p$ is the circumferential length of the combustion chamber and plenum; $S_p$ is the section of the combustion chamber and plenum; $L_i$ is the burner's length and $S_i$ its section; $\rho_p$, $\rho_c$ and $c_c$, $c_p$ are the mean-flow densities/speeds of sound of the combustion chamber and plenum, respectively (Tables \ref{tab:geo} and \ref{tab:cases}). 
The unsteady heat-release rate, $\hat{\mathcal{Q}}$, is related to the acoustic velocity at the burner's location, $\hat{u}_i$, as a Flame Transfer Function \cite{Crocco1956,Lieuwen2012}
\begin{align}\label{eq:flameresponse}
\hat{\mathcal{Q}}=\frac{\bar{p}\gamma}{\gamma-1}S_in_i\exp(\mathrm{i}\omega\tau_i)\hat{u}_i,
\end{align}
where $\gamma$ is the unburnt-gas heat capacity ratio and $\bar{p}$ is the static pressure. 
As explained by Lieuwen \cite{Lieuwen2012}, the velocity-coupled flame response model in equation \eqref{eq:flameresponse} is valid for low-frequency thermo-acoustic oscillations, where the pressure-coupled response is negligible, especially in low-Mach number flows. The response to equivalence ratio perturbations is assumed negligible. Although beyond the scope of this paper, the sensitivity method proposed here can be applied to pressure and equivalence ratio coupled flame response models without substantial modification. 

\begin{table}[!htb]
\centering 
\renewcommand{\arraystretch}{1.8} % Default value: 1
{\small{\begin{tabular}{c | c | c | c | c | c | c | c | c | c  }
$L_i$ ($m$)& $S_p$ ($m^2$) & $S_i$ ($m^2$)& $\rho_p$ ($\frac{kg}{m^3}$)& $\rho_c$ ($\frac{kg}{m^3}$) & $c_p$ ($\frac{m}{s}$)& $c_c$ ($\frac{m}{s}$) & $\alpha$ & $L_p$ ($m$) & Burners \\
\Cline{2pt}{1-10}
0.0551 & 0.00785  &  0.00028 & 10.93 & 4.7 & 519 & 832 & 0.862 & 0.54 & 19 \\
\end{tabular}}}\caption{Geometric and flow parameters of the annular combustors analysed. Same parameters as \cite{Bauerheim2016}.
%Burner length, $L_i$; 
%plenum and combustion chamber section, $S_p$; 
%burner section, $S_i$;
%plenum mean-flow density, $\rho_p$;
%combustion chamber mean-flow density, $\rho_c$;
%plenum mean-flow speed of sound, $c_p$;
%combustion chamber mean-flow speed of sound, $c_c$;
%adimensional flame position, $\alpha$;
%plenum and combustion chamber length,  $L_p$.
}\label{tab:geo}
\end{table}

\begin{figure}[!htb]
  \begin{center} 
 \includegraphics[scale=0.5]{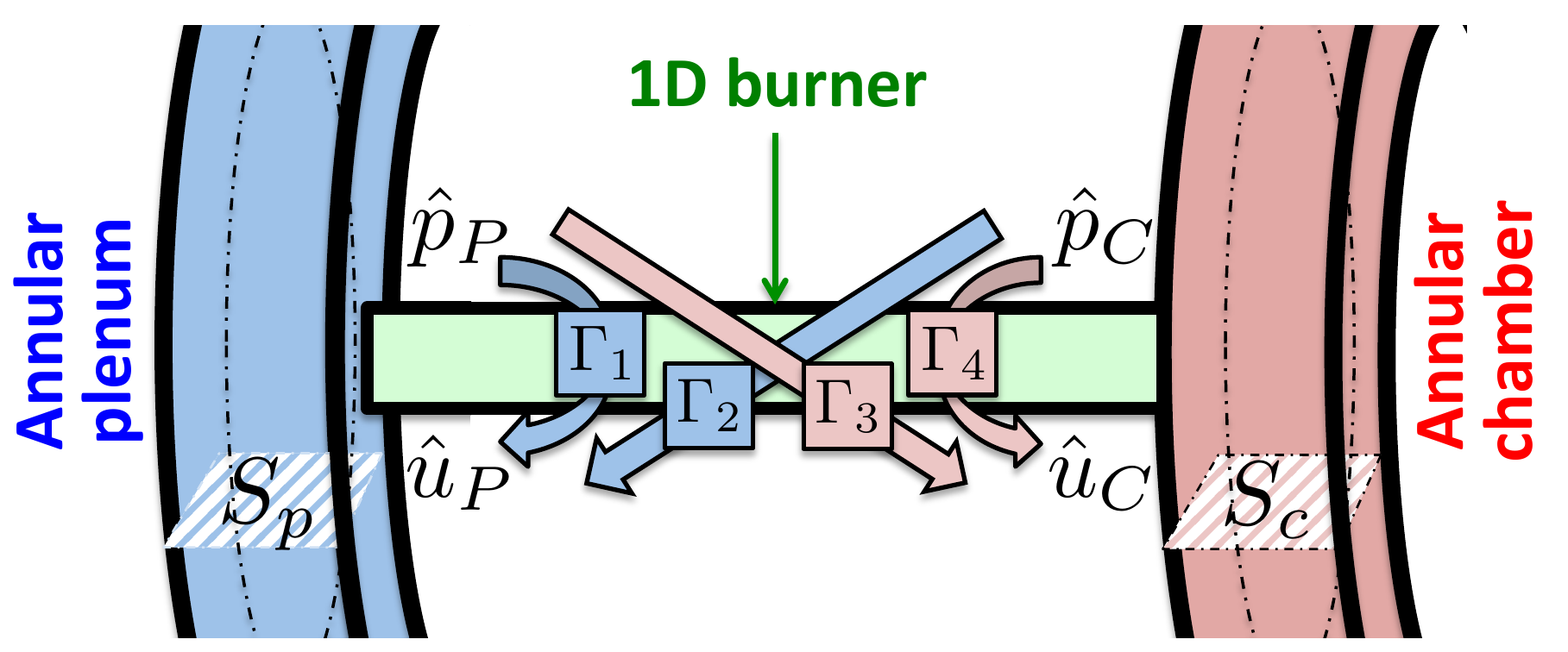}
  \end{center}
  \caption{H-connector including the burner and the flame, which connects the annular plenum to the combustion chamber. This element is analytically modelled with four coefficients $\Gamma_{ik}$, where $k=1,..,4$.}\label{fig:H_coupling}
\end{figure}

The transfer operator $\calT_i\in\mathbb{C}^{4\times4}$ maps the acoustics in the H-connector (Fig.~\ref{fig:H_coupling}), taking into account the cross-interaction of the annular plenum and chamber via the burners and the flames
\begin{align}
 \calT_i=\mathcal{I}+
 2\left( 
 \begin{matrix} 
       0 & 0 & 0 & 0\\
      \Gamma_{i1} & 0 & \Gamma_{i2} & 0\\
      0 & 0 & 0 & 0\\
      \Gamma_{i3} & 0 & \Gamma_{i4} & 0
   \end{matrix}
   \right). 
\end{align}
The coupling factors $\Gamma_{ij}$ contain the flame transfer functions and depend on geometric and flow properties. The detailed expressions are given by equations (16)--(21) in \cite{Bauerheim2014a}.
The eigenfunction is given by the state
\begin{align}
&\hat{\vecq}=\left( \begin{matrix} 
      \hat{p}_p\\
      -\mathrm{i}\rho_pc_p \hat{u}_p \\
            \hat{p}_c\\
      -\mathrm{i}\rho_cc_c \hat{u}_c \\
   \end{matrix}\right),
   \end{align}
   where $\textrm{i}^2=-1$; $ \hat{p}$ is the acoustic pressure and $ \hat{u}$ is the acoustic velocity in the two annular cavities at one particular azimuthal location. Note that knowing the state $\hat{\vecq}$ at one location is sufficient to re-build the complete acoustic field in the annular system through the propagation and transfer matrices, $\calR_i$ and $\calT_i$. 
\section{Sensitivity analysis}\label{sec:adjalgo}
The eigenvalue of interest is calculated by solving for the nonlinear characteristic equation \eqref{eq:det}. We use a Levenberg-Marquardt algorithm with the initial relevant parameter set to 0.01. (We found that trust-region-dogleg and trust-region-reflective algorithms are less accurate for this problem.) The initial guess of the frequency is the natural frequency of the combustion chamber. The numerically converged eigenvalue is denoted $\omega_0$. 

To calculate the first- and second-order sensitivities of the thermo-acoustic nonlinear eigenproblem, we implemented the following algorithm. 
\begin{enumerate}
\item We numerically evaluate the operator matrix $\calN$ substituting $\omega_0$ into equation \eqref{eq:mod1}.
\item We solve for the eigenfunction $\hat{\vecq}_0$ by Singular Value Decomposition of \eqref{eq:mod1}. The kernel is given by the singular vector associated with the null singular value. In the case of 2-fold degeneracy in rotationally symmetric annular combustors, the non-trivial kernel consists of two vectors. 
%This operator introduces a little numerical error, in that we found that $\calN\vecq\approx O(10^{-6}-10^{-8})$. This poses only a limit on the choice of $\epsilon$, which is not to be smaller than this error. 
%
\item We solve for the adjoint eigenfunction $\hat{\vecq}_0^+$ by Singular Value Decomposition of the Hermitian of \eqref{eq:mod1}, in a similar way to the previous point. 
\item By choosing a small perturbation parameter $\epsilon$, we numerically calculate the perturbation matrices for each system's parameter, $p_i$, 
$\delta_p\calN=\calN\{\omega_0,(1+\epsilon)p_i\}-\calN\{\omega_0,p_i\}$. 
\item  The operator derivative $\partial\calN/\partial\omega$  is approximated by finite difference. 
In this problem, we choose a second-order central scheme but the derivative can be evaluated analytically as well. 
\item As for the first-order sensitivities, if the problem is degenerate, as in rotationally symmetric annular combustors, we solve for the eigenproblem \eqref{eq:degenerate}. The solution consists of two eigenpairs $\{\omega_1,(\alpha_1,\alpha_2)\}$, which provide the first-order eigenvalue drift and directions caused by the perturbation to the thermo-acoustic parameters, $\vecp$. The two directions $(\alpha_1, \alpha_2)_1$ and $(\alpha_1, \alpha_2)_2$  correspond to the splitting of the eigenvalue due to symmetry breaking, assuming that the perturbation is not rotationally symmetric \cite{Bauerheim2016}. We select the greatest eigenvalue drift because we are interested in the greatest change in stability. If the problem is non degenerate, we use  equation \eqref{2eq:eigpert5}, which gives directly the first-order eigenvalue drift. 
\item We calculate the eigenvector drift, $\hat{\vecq}_1$, with the Singular Value Decomposition detailed in Section \ref{sec:PerEigf}. 
\item We calculate the second-order sensitivities with equation \eqref{2eq:eigpert8} regardless of the rotational symmetry. 
%
%\item The perturbed eigenfunction $\vecq_1$ is calculated with one of the methods explained in section \ref{} and the sec
\item For sake of comparison, the sensitivities are normalized as $\omega_{1,n}=p_i\partial\omega/\partial p_i$ and $\omega_{2,n}=1/2p_i^2\partial^2\omega/\partial p_i^2$ where $i$ denotes the $i$-th parameter. Hence, the first- and second-order eigenvalue drifts are obtained by multiplying the corresponding normalized sensitivities by $\epsilon$ and $\epsilon^2$, respectively. 
\end{enumerate}
\subsection{Validation of the adjoint gradients}\label{sec:results}
The first step is to validate the sensitivities obtained by the adjoint approach. 
We study three different systems whose schematic is depicted in Fig. \ref{fig:network}. Table \ref{tab:geo} summarizes the common parameters and Table \ref{tab:cases} summarizes the differences. 
\begin{table}[!htb]
\centering 
\renewcommand{\arraystretch}{1.5} % Default value: 1
\small{\begin{tabular}{c | c | c | c | c | c | c}
Case & $n$ & $\tau$ ($s$)& \makecell{Rotational symmetry \\ (degeneracy)}& $\mathrm{Re}(\omega_0)$ ($rad/s$)& $\mathrm{Im}(\omega_0)$ ($1/s$) & $\omega_{1,AD}$ \\
\Cline{2pt}{1-7}
A & 0.5  & 6.223$\times10^{-4}$    & Yes & {5.059$\times10^{3}$} & {-1.392} & Eq. \eqref{eq:degenerate}\\
B & 1.75  & 7.35$\times10^{-4} $   & Yes & {4.272$\times10^{3}$} & {+3.600$\times10^{-3}$} & Eq. \eqref{eq:degenerate}\\
C & 1.773  & 7.338$\times10^{-4} $   & No & {4.304$\times10^{3}$} & {-4.529$\times10^{-4}$} & Eq. \eqref{2eq:eigpert5}
\end{tabular}}\caption{The three cases analysed with common parameters in Table \ref{tab:geo}. Flame index, $n$; flame time delay, $\tau$; angular frequency, $\mathrm{Re}(\omega_0$); growth rate, $\mathrm{Im}(\omega_0)$; first-order adjoint eigenvalue drift, $\omega_{1}$. For Case C, $n$ and $\tau$ are the mean values of the flame parameter distribution of Fig. \ref{fig:fla_par_C}. The second-order adjoint eigenvalue drift is given by equation \eqref{2eq:eigpert8} regardless of the case. The flame parameters of Cases A and B are the same as "Wcase" and "Scase", respectively, in \cite{Bauerheim2016}. Here we used a slightly smaller $\tau$ for case A to have a stable configuration.}\label{tab:cases} 
\end{table}

\begin{figure}[!htb]
    \centering
        \includegraphics[width=0.6\linewidth]{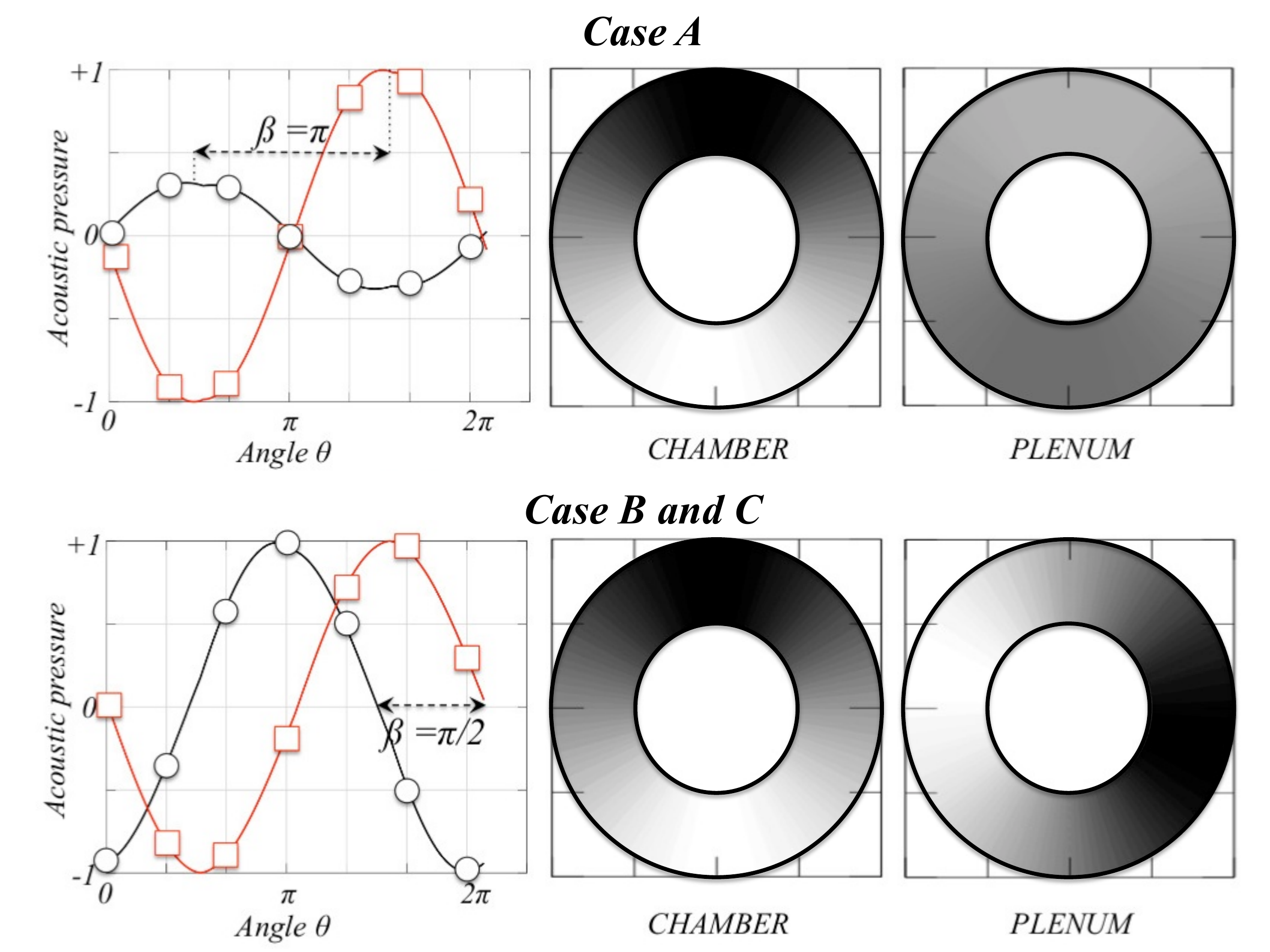}
        \caption{Left column: Acoustic pressure along the azimuthal direction in the chamber ($\square$) and plenum ($\circ$). Their relative amplitudes and phase lags,  $\beta$, are highlighted. Middle and right columns: pressure field in the combustion chamber and plenum. Top row: weakly coupled Case A. Bottom row: strongly coupled Cases B and C.}
        \label{fig:weakstrong}
\end{figure}

Case A belongs in the `weakly coupled region' of Bauerheim et al. \cite{Bauerheim2014a}. Physically, the thermo-acoustic mode occurs in the combustion chamber and is almost decoupled from the plenum dynamics (Fig.~\ref{fig:weakstrong}, top panels). Because of rotational symmetry, the eigenvalue is 2-fold degenerate. By assuming that the coupling factors are small, $\lVert\Gamma_{ik}\ll 1\rVert$, this case has an approximate analytical solution for the eigenvalue~\cite{Bauerheim2014a}.

%%%
%%%

\begin{figure}[!htb]
    \centering
        \includegraphics[width=0.9\linewidth]{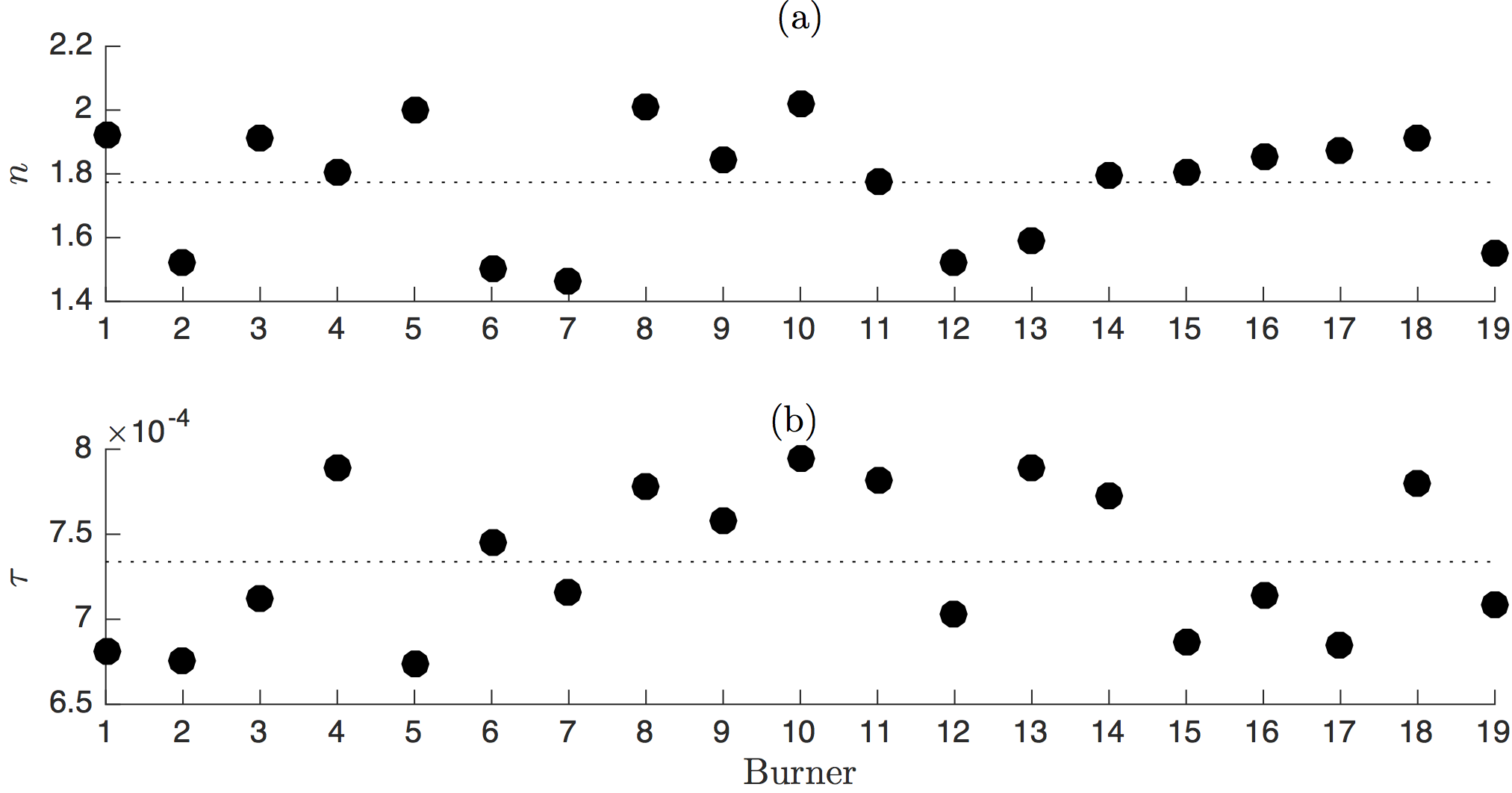}
        \caption{(a) Flame indices and (b) flame time delays of the non-rotationally symmetric Case C (non-degenerate case). The dotted lines denote the mean values.}
        \label{fig:fla_par_C}
\end{figure}

Case B belongs in the `strongly coupled region' of \cite{Bauerheim2014a}, where the thermo-acoustic mode in the combustion chamber is coupled with the mode in the plenum through longitudinal acoustic waves in the burners (Fig.~\ref{fig:weakstrong}, bottom panels). As for Case A, because of rotational symmetry, the eigenvalue is 2-fold degenerate.
Case C corresponds to Case B but the symmetry is broken by uniformly random flame parameters around their mean values (Fig. \ref{fig:fla_par_C}), therefore, the eigenvalue is not degenerate.

\begin{figure}[!htb]
    \centering
        \includegraphics[width=0.99\linewidth]{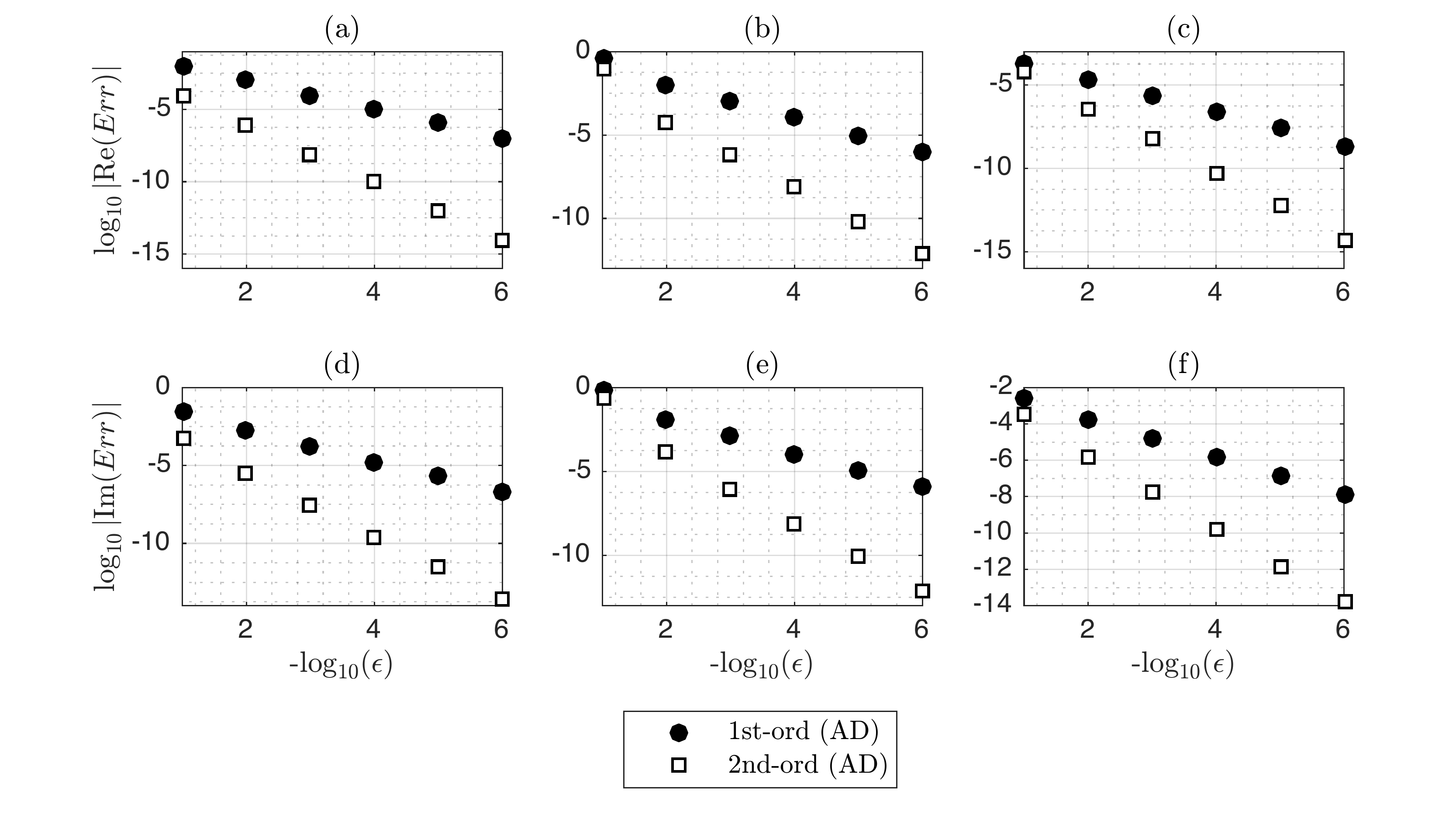}
        \caption{Relative error between finite-difference and first- and second-order adjoint sensitivities. Case A in the first column; Case B in the second column and Case C in the last column. Top panels: angular frequency relative error; bottom panels: growth-rate relative error. Sensitivities calculated by randomly perturbing the flame parameters $n$ and $\tau$. The benchmark solution is calculated by finite difference. The logarithmic slopes of $-1$ and $-2$ show that the adjoint sensitivities are correct up to first- (circles) and second-order (squares). }
        \label{ffig:err_deg}
\end{figure}

The relative error, which is a complex quantity, is defined as 
\begin{align}\label{eq:err}
Err = \frac{\omega^{AD} - \omega^{FD}}{\epsilon\lvert\omega_0\lvert},
\end{align}
where the superscript FD denotes `Finite Difference', which is the benchmark solution, and AD denotes `Adjoint'. 
The systems are perturbed by random $\epsilon$-perturbations to the flames' indices, $n$, and time delays, $\tau$. The symmetry-breaking random $\epsilon$-perturbation to the 19 $n,\tau$ parameters is inputted into equation \eqref{eq:general} and the nonlinear eigenproblem is solved to find $\omega^{FD}$. The same perturbation is then inputted into the first- and second-order adjoint equations \eqref{2eq:eigpert5},\eqref{eq:degenerate},\eqref{2eq:eigpert8}. The first-order $\omega^{AD}$ is given by $\omega_0+(\epsilon\mathbf{p}_1)\omega_1$ and the second-order by $\omega_0+(\epsilon\mathbf{p}_1)\omega_1+\frac{1}{2}(\epsilon\mathbf{p}_1)^2\omega_2$ (Fig.~\ref{ffig:err_deg}). 
% The first- and second-order are shown in Figures \ref{ffig:err_deg}, \ref{fig:err_nondeg_stro}, respectively. 
The real and imaginary parts of the error for the three cases are correct up to first- and second-order because the logarithmic slope is $-1$ and $-2$, respectively. 

% \begin{figure}
% \centering
%         \includegraphics[width=0.9\linewidth]{Error2.eps}
%         \caption{Same as Fig. \ref{ffig:err_deg} as for the second-order sensitivities. The logarithmic slope of $-2$ shows that the sensitivities are correct up to second order. }
%         \label{fig:err_nondeg_stro}
% \end{figure}
% %
%
  %
  \subsection{Eigenvalue sensitivity to thermo-acoustic design parameters}\label{sec:sens}
Because of their complexity, annular combustor network models usually contain a large number of parameters, typically tens or hundreds. 
%To understand the predicted stability based on such models with a large number of uncertain parameters, it is necessary to know the sensitivity of the quantity of interest, which, in this study, is the eigenvalue. 
A ranking of the parameters to which the eigenvalue is most sensitive informs the designer/analyst about the most critical model assumptions/parameters that influence the stability. %The information from these sensitivities can be then used to improve current models and study uncertainty quantification of the most uncertain parameters. 
Previous studies calculated the first-order sensitivities to flame parameters and external controller  in single-flame longitudinal configurations \cite[see, e.g.,][]{Magri2013}. Here, we calculate both  first- and second-order adjoint sensitivities, which are computationally cheap, and compare them with the finite-difference calculations, which involve numerous computations of nonlinear eigenproblems. All the sensitivities discussed are normalized as suggested at point 9 in Section \ref{sec:adjalgo} to ease the comparison. 

\begin{figure}[!htb]
        \centering
        \includegraphics[width=0.99\linewidth,draft=false]{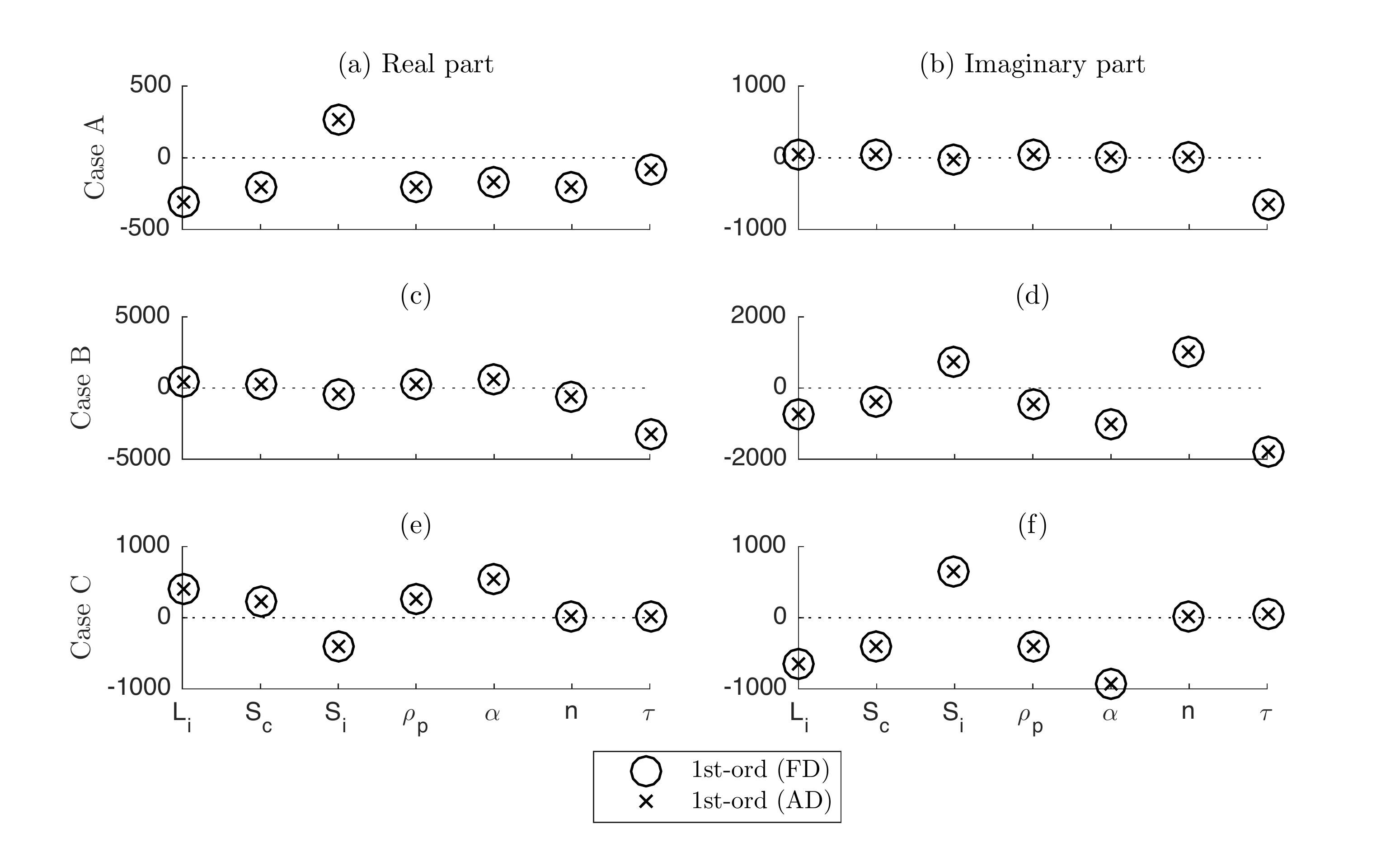}
        \caption{Normalized first-order eigenvalue sensitivities. Calculation with Finite Difference (FD) and Adjoint methods (AD). The angular frequency sensitivity is shown in the left panels, the growth-rate sensitivity is shown in the right panels. Case A in the top row; Case B in the middle row and Case C in the bottom row. The adjoint sensitivity matches the benchmark solution given by finite differences. In Case C, the sensitivity of $n$ and $\tau$ is the mean value of the single-burner sensitivities of Figs. \ref{fig:sens_flame_n} and \ref{fig:sens_flame_tau}. The first-order eigenvalue drift is obtained by multiplying these sensitivities by $\epsilon$.}
        \label{fig:sens_geo1}
\end{figure}

\begin{figure}[!htb]
        \centering
        \includegraphics[width=0.99\linewidth,draft=false]{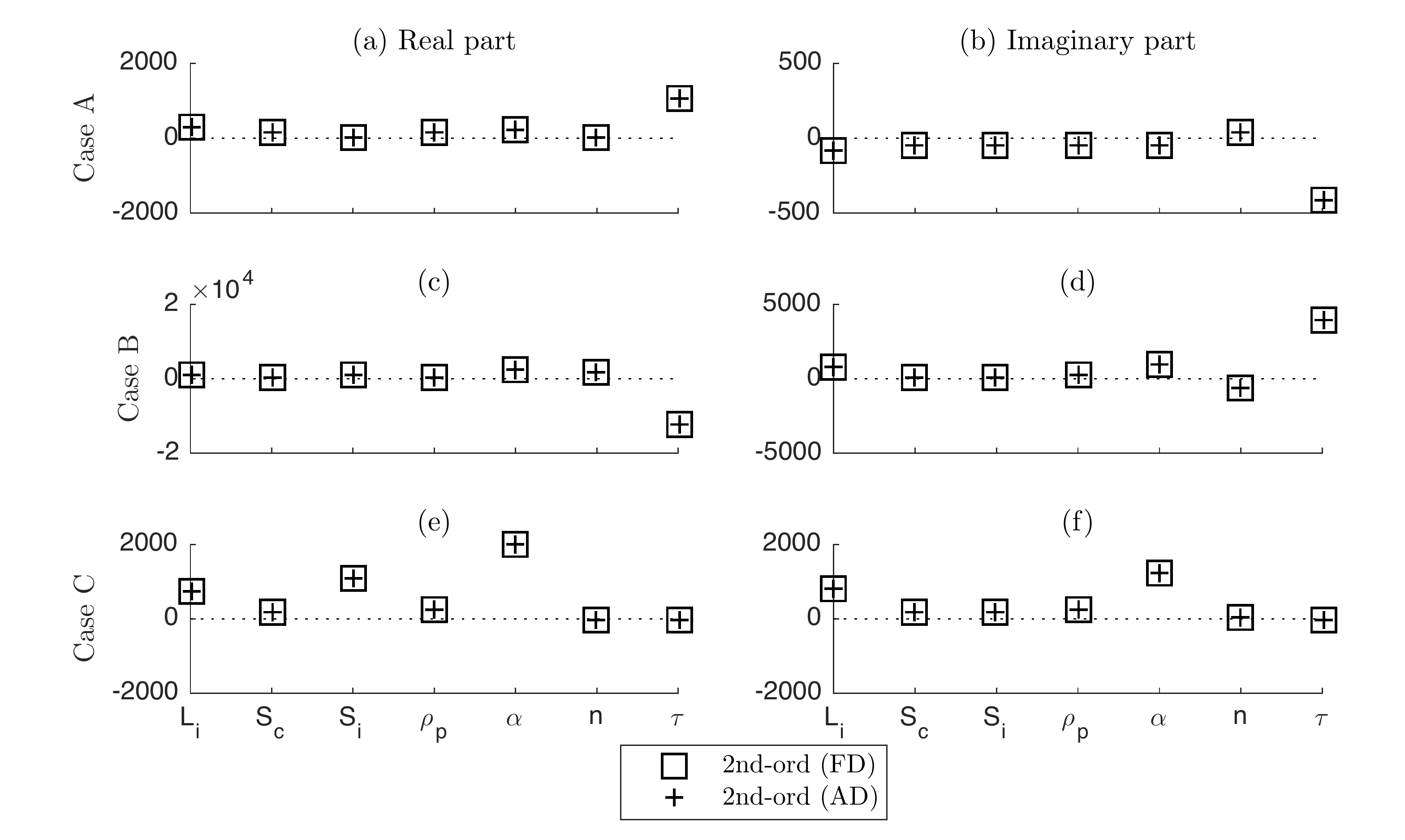}
        \caption{Same as Fig. \ref{fig:sens_geo1} but for second-order sensitivities. The normalized second-order sensitivities are higher than the first-order sensitivities of Fig.~\ref{fig:sens_geo1}. The second-order eigenvalue drift is, however, smaller because it is obtained by multiplying these sensitivities by $\epsilon^2$.}
        \label{fig:sens_geo2}
\end{figure}

In Fig. \ref{fig:sens_geo1} the first-order sensitivities are presented for the three cases. 
%The real part is the angular-frequency sensitivity (left column), whereas the imaginary part is the growth-rate sensitivity (right column). 
In Cases A and B, only one flame (modelled with two parameters, the gain $n$ and the time-delay $\tau$) is perturbed because the configuration is axi-symmetric. Consequently, the FD method requires solving $5+1+1=7$ nonlinear eigenproblems, as many as the number of parameters under investigation, imposing a relative perturbation of $\epsilon=10^{-6}$. (This is sufficiently small to accurately estimate the gradients by first-order forward finite difference. Other choices of $\epsilon$ were tested and no significant difference was found).  
Via the adjoint approach, only one nonlinear eigenproblem and its adjoint have to be solved. 
In Case C, each flame is perturbed independently because the configuration in not axi-symmetric. This means that the FD approach requires solving for $5+19+19=43$ nonlinear eigenproblems, whereas the adjoint formulation still requires only one eigenproblem and its adjoint to be calculated. 
Figure~\ref{fig:sens_geo2} shows the second-order sensitivities for the same quantities. 
The adjoint predictions match the benchmark solutions given by finite differences. The sensitivities are reported in Table \ref{tab:sens}. 

\begin{figure}[!htb]
        \centering
        \includegraphics[width=0.99\linewidth,draft=false]{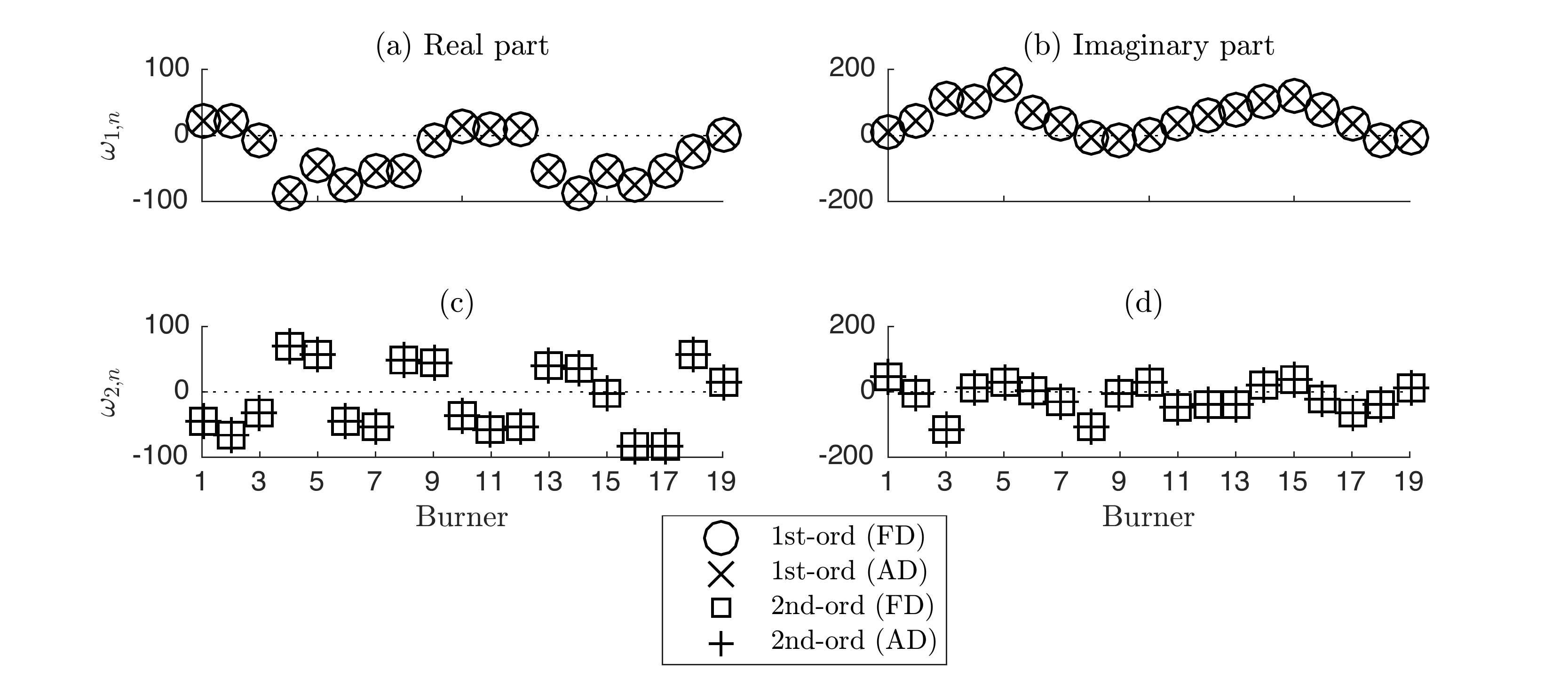}
        \caption{First- (top row) and second-order (bottom row) sensitivities to the flame index, $n$, in Case C. The sensitivities vary because the configuration is non-rotationally symmetric (Fig. \ref{fig:fla_par_C}). Angular frequency sensitivity in the left panels, growth-rate sensitivity in the right panels. The adjoint sensitivity matches the benchmark solution given by finite differences. The first-order eigenvalue drift is obtained by multiplying these sensitivities by $\epsilon$.}
        \label{fig:sens_flame_n}
\end{figure}
\begin{figure}[!htb]
        \centering
        \includegraphics[width=1.0\linewidth,draft=false]{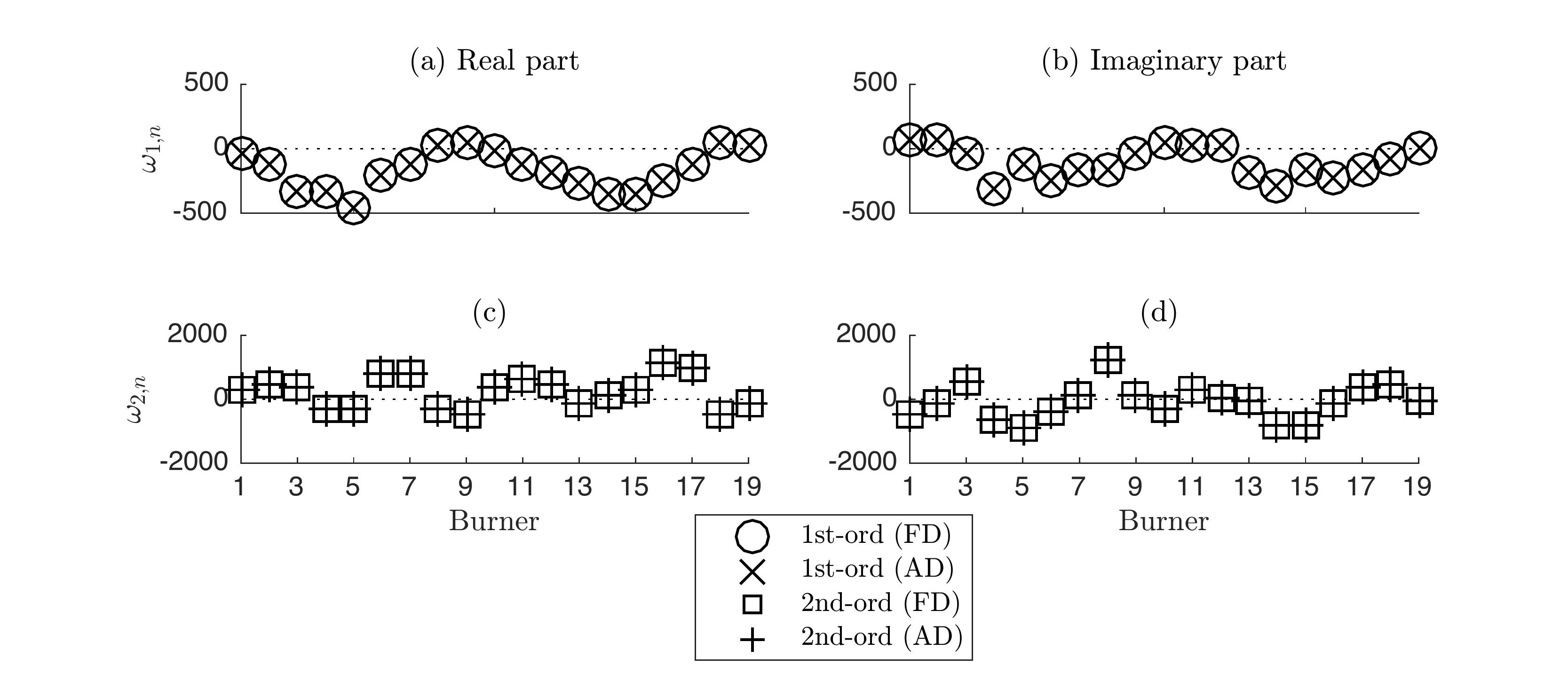}
        \caption{Same as Fig. \ref{fig:sens_flame_n} as for the sensitivity to the time delay, $\tau$. The second-order eigenvalue drift is obtained by multiplying these sensitivities by $\epsilon^2$.}
        \label{fig:sens_flame_tau}
\end{figure}

 \begin{table}[!htb]
\centering
\setlength{\tabcolsep}{10pt} % Default value: 6pt
\renewcommand{\arraystretch}{1.2} % Default value: 1
  {\small{\begin{tabular}{ c |c | c | l | l |  l | l | l | l | l  }
& & Case & $L_i$ & $S_p$ & $S_i$ & $\rho_p$ & $\alpha$ & $n$& $\tau$ \\ \Cline{2pt}{1-10}
  \parbox[t]{2mm}{\multirow{6}{*}{\rotatebox[origin=c]{90}{1st-order}}} &   \parbox[t]{2mm}{\multirow{3}{*}{\rotatebox[origin=c]{90}{$\mathrm{Re}(\omega_{1,n})$}}}   &  
          A & xxx & xx&xxx & xx & xx & xx & x \\
      & & B & x & x&x & x & x & x & xxx \\
      & & C & xx & x&xx & x & xx & x & x \\  \Cline{1pt}{2-10}
     & \parbox[t]{2mm}{\multirow{3}{*}{\rotatebox[origin=c]{90}{$\mathrm{Im}(\omega_{1,n})$}}} &   A & x& x&x & x & x & x & xxx \\
     &  & B & xx& x&xx & x & xx & xx & xxx \\
     & & C & xxx & xx&xxx & xx & xxx & x & x \\ 
     \Cline{1pt}{1-10}
  \parbox[t]{2mm}{\multirow{6}{*}{\rotatebox[origin=c]{90}{2nd-order}}} &   \parbox[t]{2mm}{\multirow{3}{*}{\rotatebox[origin=c]{90}{$\mathrm{Re}(\omega_{2,n})$}}}   &  
          A & x & x&x & x & x & x & xxx \\
      & & B & x & x&x & x & x & x & xxx \\
      & & C & xx & x&xx & x & xxx & x & x \\  \Cline{1pt}{2-10}
         &    \parbox[t]{2mm}{\multirow{3}{*}{\rotatebox[origin=c]{90}{$\mathrm{Im}(\omega_{2,n})$}}}   &  
          A & x& x&x & x & x & x & xxx \\
     &  & B & x& x&x & x & x & x & xxx \\
     & & C & xx & x&x & x & xxx & x & x \\ 
  \end{tabular}}}
\caption{Summary of the sensitivities of Figs. \ref{fig:sens_geo1},\ref{fig:sens_geo2}, \ref{fig:sens_flame_n} and \ref{fig:sens_flame_tau}. xxx=strong, xx=mild, x=weak.}\label{tab:sens} 
\end{table}
In the rotationally symmetric Case A (Fig. \ref{fig:sens_geo1}a), the dominant parameters governing the  first-order frequency sensitivity are geometric, being the burner length, $L_i$, and the burner cross-sectional area, $S_i$.  On the other hand, the  first-order growth-rate sensitivity is highest in the flames' time delays $\tau$ (Fig. \ref{fig:sens_geo1}b). 
This high time-delay sensitivity was already shown in previous adjoint-based studies in simpler geometries \cite{Magri2013}.
The flame index $n$ plays a minor role in the stability sensitivity in configuration A. 
Whereas the system is still most sensitive to second-order perturbations in the time delay (Fig. \ref{fig:sens_geo2}b), the frequency is less affected by perturbations in the geometry (Fig. \ref{fig:sens_geo2}a). On the other hand, Case B displays similar sensitivities to second-order perturbations, with a more pronounced dependence on the flame index (Fig. \ref{fig:sens_geo2}d). 
For both rotationally-symmetric cases, the second-order sensitivities are highest in the time delay. However, the strongly coupled system B is markedly more sensitive  to the flame parameters than system A (Figs. \ref{fig:sens_geo1},\ref{fig:sens_geo2}a,b,c,d). 
The difference between the sensitivity of Cases A and B suggests that when the plenum resonance is not coupled with the combustion chamber modes through the longitudinal modes (Case A), the overall stability is more susceptible to small changes in the geometry rather than the flame parameters. 

The non-rotationally symmetric strongly coupled case (Case C) shows a different behaviour. 
The system is significantly less affected by perturbations in the flame parameters, as shown in Fig. \ref{fig:sens_geo1}e,f and Fig. \ref{fig:sens_geo2}e,f. (Here, the sensitivities of $n$ and $\tau$ are the mean values of the single-burner sensitivities of Figs. \ref{fig:sens_flame_n} and \ref{fig:sens_flame_tau}). The system is sensitive to changes in the geometry, especially in the burner section and flame position.  These observations hold for both first- and second-order sensitivities of the frequency and growth rate. 
Because Case C has different flames, the flame sensitivity changes from burner to burner, as shown in Figs. \ref{fig:sens_flame_n} and \ref{fig:sens_flame_tau}. (Their mean values are shown in Fig. \ref{fig:sens_geo1}e,f and Fig. \ref{fig:sens_geo2}e,f.) The thermo-acoustic system has drastically different behaviours depending on the burner being perturbed. The first-order sensitivities oscillate in the azimuthal direction and can be negative or positive. This $2-$periodic pattern  is physically related to the eigenvalue splitting caused by symmetry breaking~\cite{Taylor2011,Bauerheim2015}, which is due to the $2$-nd Fourier coefficient of the flame parameters' spatial distribution ($C_{2n}$ in \cite{Taylor2011}). From Figs.~\ref{fig:sens_flame_n}c,d and \ref{fig:sens_flame_tau}c,d, we note that the second-order sensitivity patterns are $4-$periodic. This oscillation might be due to the $4$-th Fourier coefficient of the flame distribution. This analysis, however, is beyond the scope of this paper and is left for a follow-on study. 
By inspection, we find an accurate match between the finite difference calculations and the adjoint predictions. 
The growth rate is overall most sensitive to the time delays $\tau$, but the value is about twenty times smaller than the corresponding rotationally symmetric Case B of Fig.~\ref{fig:sens_geo1}d. Moreover, although configuration C is similar to the corresponding rotationally symmetric Case B, the parameters to which it is most sensitive are different. This might indicate that modelling an annular combustor as a rotationally symmetric configuration might overestimate and poorly predict the sensitivities. 
%
%
%
%
%
  %
%%%%%%%%%%%%%%%%%%%%%%%%%%%%%%%
%%%%%%%%%%%%%%%%%%%%%%%%%%%%%%%
% CONCLUSIONS  
%%%%%%%%%%%%%%%%%%%%%%%%%%%%%%%
%%%%%%%%%%%%%%%%%%%%%%%%%%%%%%%
\section{Conclusions} 
We present first- and second-order sensitivities of eigenvalues in nonlinear non-self-adjoint eigenproblems with/without degeneracy via an adjoint method. 
This is the first application of adjoint sensitivity analysis to nonlinear eigenproblems as applied to design-parameter sensitivity studies in thermo-acoustics. 
The adjoint sensitivities are calculated in an elaborate annular combustor thermo-acoustic network. 
Two cases are studied as representative cases of plenum-combustion-chamber dynamics: the weakly coupled case, in which the combustion-chamber mode is unstable, and the strongly coupled case, in which the plenum mode is coupled with the combustion-chamber mode through the burners. 

We show how to use the adjoint framework to study the sensitivity to the system's parameters reducing the number of computations by a factor equal to the number of the system's parameters. This is particularly attractive to annular combustors, where the number of flames, thus parameters, is large. We find the strongly coupled case is overall more sensitive and the symmetry breaking makes the system less sensitive. This suggests that perfect rotational symmetry might be an exceedingly sensitive model. Moreover, the adjoint sensitivities are not prone to numerical cancellation errors, in contrast to finite differences, because they do not depend on the size of the perturbation. 

The sensitivity analysis showed that the eigenvalue can be most sensitive to geometric parameters (see cases A and C in Table \ref{tab:sens}). However, the manufacturing tolerances on the geometry are usually small, i.e., the uncertainty on the physical dimensions of the annular combustor is small. On the other hand, the uncertainty on the flame parameters and/or damping is larger \cite{OConnor2015}. In order to evaluate the probability that a thermo-acoustic mode is unstable, the adjoint method proposed is extended to uncertainty quantification of the eigenvalue calculation in the second part of this paper \cite{Magri2016jcp2}. 

The adjoint framework is a promising method for design to obtain quick estimates of the thermo-acoustic  sensitivities  at very cheap computational cost. 

%%%%%%%%%%%%%%%%%%%%%%%%%%%%%%%
%===================================================================
\subsection*{Acknowledgments}
%===================================================================

The authors are grateful to the 2014 Center for Turbulence Research Summer Program (Stanford University) where the ideas of this work were born. 
L.M. and M.P.J acknowledge the European Research Council - Project ALORS 2590620 for financial support. L.M gratefully acknowledges the financial support received from the Royal Academy of Engineering Research Fellowships scheme. The authors thank Prof. Franck Nicoud for fruitful discussions. Figure \ref{fig:adjnet} was adapted from the article of S. R. Stow and A. P. Dowling, {\it A time-domain network model for nonlinear thermoacoustic oscillations}, ASME Turbo Expo, GT2008-50770 \cite{Stow2008} with permission of the original publisher ASME.
%%
%%%%%%%%%%%%%%%%%%%%%%%%%%%%%%%
%%%%%%%%%%%%%%%%%%%%%%%%%%%%%%%
% APPENDIX 
%%%%%%%%%%%%%%%%%%%%%%%%%%%%%%%
%%%%%%%%%%%%%%%%%%%%%%%%%%%%%%%
\section*{Appendix A. Restricted matrix inversion for the calculation of the perturbed eigenfunction}\label{app:q1}
%\subsection*{A.1. Restricted matrix inversion}
%
The aim is to find the square submatrix of $\calN\{\omega_0,\mathbf{p}_0\}$ with rank = $K-N$, in the subspace of which the matrix is invertible. 
First, we partition $\calN\hat{\vecq}_0$ as 
\begin{align}\label{eq:nels}
\calN\{\omega_0,\mathbf{p}_0\}\hat{\vecq}_0=\left( \begin{matrix} 
      \mathbf{N}_{11} & \mathbf{N}_{1k} &\mathbf{N}_{13}\\
      \mathbf{N}_{k1} & \mathbf{N}_{kk} &\mathbf{N}_{k3}\\
      \mathbf{N}_{31} & \mathbf{N}_{3k} &\mathbf{N}_{33}
   \end{matrix}\right)
   \left( \begin{matrix} 
      \hat{\vecq}_{0,1} \\
      \hat{\vecq}_{0,k} \\
      \hat{\vecq}_{0,3} 
   \end{matrix}\right), 
\end{align}
where $\mathbf{N}_{1k}$, $\mathbf{N}_{3k}$ are $K\times N$ submatrices; $\mathbf{N}_{k1}$, $\mathbf{N}_{k3}$ are $N\times K$ submatrices and $\mathbf{N}_{kk}$ is an $N\times N$ submatrix. 
$ \hat{\vecq}_{0,1}, \hat{\vecq}_{0,k}, \hat{\vecq}_{0,3}$ are subvectors. 
The subvector $\hat{\vecq}_{0,k}$ is chosen to have non-trivial components. Hence, the system can be recast as
\begin{align}\label{eq:nels2}
\calN\{\omega_0,\mathbf{p}_0\}\hat{\vecq}_{0}=
\left( \begin{matrix} 
      \mathbf{N}_{11} & \mathbf{N}_{13}\\
      \mathbf{N}_{k1} &\mathbf{N}_{k3}\\
      \mathbf{N}_{31}  &\mathbf{N}_{33}
   \end{matrix}\right)
   \left( \begin{matrix} 
      \hat{\vecq}_{0,1} \\
      \hat{\vecq}_{0,3} 
   \end{matrix}\right) 
   = 
   - \left( \begin{matrix} 
      \mathbf{N}_{1k} \\
      \mathbf{N}_{kk} \\
      \mathbf{N}_{3k} 
   \end{matrix}\right)
    \begin{matrix} 
      \hat{\vecq}_{0,k}. 
         \end{matrix}
\end{align}
The matrix on the left-hand side has rank = $K-N$ because all its columns are independent since the $N$ values of $\hat{\vecq}_{0,k}$ are non-trivial and the right-hand side is a linear combination of the left-hand side.  Therefore, the columns of $\calN$ corresponding to $\hat{\vecq}_{0,k}$ on the left hand-side can be removed from the matrix without affecting its rank. 
To reduce the row space to a subspace in which the matrix has full rank, we use the same argument as before with the adjoint eigenvector $\hat{\vecq}_{0}^+$. The $N$ components $\hat{\vecq}_{0,k}^+$ are chosen to be non-trivial. Hence, the $N$ rows corresponding to $\hat{\vecq}_{0,k}^+$ can be removed and the final linear system becomes invertible in this subspace, as follows 
\begin{align}\label{eq:nels3}
\left( \begin{matrix} 
      \mathbf{N}_{11} & \mathbf{N}_{13}\\
      \mathbf{N}_{31}  &\mathbf{N}_{33}
   \end{matrix}\right)
   \left( \begin{matrix} 
      \hat{\vecq}_{0,1} \\
      \hat{\vecq}_{0,3} 
   \end{matrix}\right) 
   = 
   - \left( \begin{matrix} 
         \mathbf{N}_{1k} \\
      \mathbf{N}_{3k} 
   \end{matrix}\right)
    \begin{matrix} 
      \hat{\vecq}_{0,k}. 
         \end{matrix}
\end{align}
Now, the square matrix on the left-hand side is invertible because it has rank = $K-N$ and the subspace dimension is $K-N$. 
Using this observation, we can solve for the perturbed eigenvector substituting equation \eqref{eq:nelsdec} into \eqref{eq:nels3}
\begin{align}\label{eq:nels4}
\left( \begin{matrix} 
      \mathbf{N}_{11} & \mathbf{N}_{13}\\
      \mathbf{N}_{31}  &\mathbf{N}_{33}
   \end{matrix}\right)
   \left( \begin{matrix} 
      \hat{\mathbf{z}}_{1} \\
      \hat{\mathbf{z}}_{3} 
   \end{matrix}\right) 
   = 
   - \left( \begin{matrix}
      \mathbf{N}_{1k} \\
      \mathbf{N}_{3k} 
   \end{matrix}\right)
    \begin{matrix} 
      \hat{\vecq}_{0,k}
         \end{matrix}
            + \left( \begin{matrix}
      \mathbf{\Psi}_{1k} \\
      \mathbf{\Psi}_{3k} 
   \end{matrix}\right).
\end{align}
Setting the eigenvector to zero because it is already known, $\hat{\mathbf{z}}$ can be easily found by solving the linear system, the final solution of which is 
\begin{align}
\hat{\vecq}_1 =             
 \left( \begin{matrix} 
      \hat{\mathbf{z}}_{1} \\
      \mathbf{0}\\
      \hat{\mathbf{z}}_{3} 
   \end{matrix}\right)
         + 
      \beta_0\hat{\vecq}_0, 
\end{align}
where $\mathbf{0}$ is a null $N\times1$ vector. This extends the method proposed by \cite{Nelson1976} to degenerate nonlinear eigenproblems. 
%
%
%
%\subsection{A.2. Eigenexpansion}
%%
%%
%If a complete eigenbasis $(\hat{\vecq}_0,\hat{\mathbf{e}}_i)$ is known, we seek a solution of the form 
%%
%\begin{align}\label{eq:pereig}
%\tilde{{\mathbf{z}}} = \sum_{i=1}^{K-N}\beta_i\hat{\mathbf{e}}_i.
%\end{align}
%%
%Hence, $\tilde{{\mathbf{z}}}\in\mathbb{C}^{K-N}$ and the vector $\hat{\mathbf{z}}$ is $\tilde{\mathbf{z}}$ concatenated with $N$ zeros. 
%% The coefficient $\beta_0\in\mathbb{C}$ is undetermined because all the solutions $\hat{\vecq}_1=\beta_0\hat{\vecq}_0$ satisfy eq.  \eqref{2eq:eigpert3}. 
%The coefficients $\beta_i\in\mathbb{C}$ can be calculated by projecting equation \eqref{2eq:eigpert3} onto the non-degenerate adjoint eigenfunctions, $\hat{\vecq}_{0,i}^+$ (any other complete basis would suit).  This yields 
%\begin{align}\label{eq:pereig2}
%& \sum_{i=1}^{K-N}\beta_i \hat{\vecq}^+_{0,j}\cdot\calN\{\omega_0,\mathbf{p}_0\}\hat{\mathbf{e}}_i= -\hat{\vecq}^+_{0,j}\cdot \mathbf{\Psi}, \\
%& j=1,2,\ldots,K-N. 
%\end{align}
%
%%%%%%%%%%%%%%%%%%%%%%%%%%%%%%%

%



\section*{References}
                 \bibliographystyle{elsarticle-num} 
                 \bibliography{MyLibrary_2}

\end{document}